\documentclass[aps,pre,showpacs,twocolumn,groupedaddress,floatfix]{revtex4-1}

  \usepackage[latin1]{inputenc}
  \usepackage{epsfig}
  \usepackage{amssymb}
  \usepackage[dvipsnames]{xcolor}
\begin{document}

\title{Interaction of Wave-Driven Particles with Slit Structures}

\author{Clive Ellegaard and Mogens T. Levinsen}
\email{levinsen@nbi.dk}

\affiliation{Niels Bohr Institute, University of Copenhagen,
  Blegdamsvej 17,
  DK-2100 Copenhagen {\O},
  Denmark}

\date{\today}

\begin{abstract}
Just over a decade ago Couder and Fort \cite{CouderFort} published a provocative paper suggesting that a classical system might be able to simulate the truly fundamental quantum mechanical single- and double-slit experiment. The system they investigated was that of an oil droplet walking on a vibrated oil surface. Their results have since been challenged \cite{Andersen,Andersen2} by pointing to insufficient statistical support and a lack of experimental control over critical parameters. Here we show that the randomness in the original experiment is an artifact of lack of control. We present experimental data from an extensive scan of the parameter space of the system including the use of different size slits and tight control of critical parameters. For the single-slit we find very diverse samples of interference-like patterns but all causal by nature. This also holds for the double-slit. However, an extra interference effect appears here. The origin of this is investigated by blocking either the inlet or the outlet of one slit. Hereby we show that the extra interference is solely due to back-scatter of the associated wave field from the outlet of the slit not passed by the droplet. Recently Pucci {\it et al.} \cite{BushNew} using a much broader slit also showed that the classical system is basically causal. They, too, observed the extra interference effect for the double-slit. However, the reason behind was {\it not} determined.  Moreover they claimed the existence of a chaotic regime just below the critical acceleration for spontaneous generation of Faraday surface waves.  Our measurements do not support the validity of this claim. However, the drop dynamics turns out to have an interesting multifaceted interaction with the slit structure.
\end{abstract}

\maketitle

\section{Introduction:}
Many people have wondered about what lies behind quantum physics - like de Broglie who introduced the pilot-wave theory - instead of just accepting the randomness of the Copenhagen school. 

Could quantum mechanics be realistically simulated by classical experiments? That was the implicit question raised by Couder and Fort a decade ago in a letter published in PRL \cite{CouderFort} about an experimental investigation of a classical single- and double-slit experiment. The system under investigation was a combined particle-wave system consisting of an oil droplet surfing on a self-generated capillary wave field on a vibrating surface of oil, driven vertically just below the Faraday instability \cite{Couder1, Faraday}. The conclusion of the authors was that the classical particle-wave system seemingly obeyed the same statistics as a quantum system. Many quantum-like phenomena can actually be simulated with this system e.g. tunneling \cite{tunneling} and quantized orbits \cite{QuantizedOrbits}.

Besides the analogs, the system can also be used to visualize many solid-state systems as e.g. two-dimensional small crystalline structures. A beautiful visualization of Cooper pairs bound by the exchange of acoustic phonons \cite{Cooper,Couder2} is found where two droplets rotate around each other interacting through their wave fields in a bound state.

In a quantum double-slit experiment the probability wave function always propagates as a single inseparable object, and any attempt to determine which slit is used by the entity under investigation will immediately quench the self-interference of the wave function after passage. In the classical case, however, the particle will be observed to move through one distinct slit while any possible part of the wave field passing through the other slit becomes effectively decoupled from the particle. The latter wave field will therefore be decaying. In the quantum case the interference is global with linear superposition of the fields while in the case of the droplet the interference can at best be local. 

The conclusions by Couder and Fort \cite{CouderFort} have already been  challenged on theoretical grounds by Andersen {\it et al.} \cite{Andersen,Andersen2} who furthermore pointed to insufficient statistical support.  Andersen {\it et al.} \cite{Andersen} reproduced the experiment with considerably improved statistics using a setup similar to that of Couder and Fort. They also pointed out that the measured transit time through the slit was of the order of the observed wave decay time, thus making it doubtful that waves passing the unused slit could survive long enough to influence the outcome. A similar experiment was made by Batelaan {\it et al.} \cite{batelaan}. In a later experiment Andersen {\it et al.} \cite{Andersen3} measured the wave-fields generated by the droplet and found no evidence for waves passing through the slit not used by the droplet. However, most important for this investigation, a clarification about the cause of the observed interference was not reached in any of these experiments.

In a recent paper by Pucci {\it et al.} \cite{BushNew} (see also Dan Harris \cite{Harris}) it is shown, that for broad slits corresponding to the single-slit size used by Couder and Fort \cite{CouderFort} and drive amplitudes not too close to the critical amplitude for excitation of the Faraday instability, the angular deviation in the single-slit case had a causal relation to the impact parameter. They moreover investigated the two-slit situation using the same slit size and here, too, found a causal relation but also that a distinct interference effect was present since the distribution of output angles for the individual slits were symmetrically skewed with respect to the single-slit distribution. The cause of the interference was, however, not determined. Closer to the critical drive amplitude they observed  that the direction taken by the droplet after the passage of a slit evidently could be chosen chaotically, and they presented a probability density plot (PDF plot) with a reasonable statistic (235 tracks, single-slit, see their Fig. 5) in this case showing some features akin to that shown by Couder and Fort \cite{CouderFort}.

We here present results from narrow single- and double-slit experiments that mimic Couder and Fort's double-slit configuration. However, we have enforced tight control of whatever parameters are found critical. We find clear effects of interference resulting in different, but always sharp and reproducible, angular distributions of tracks between single- and double-slit experiments. This is in contrast to the random angular deflection found by Couder and Fort \cite{CouderFort}. Moreover, we present evidence that these differences are caused by back-scattering of waves from the outlet side of the slit not used by the droplet and consequently are not due to any waves passing through that slit. Lastly, these narrow slits result in much more complicated and fascinating structures making them worth a study by themselves though it is not quantum physics.   

We also look into some of the other issues raised by Pucci {\it et al.}  \cite{BushNew}, here extending our investigation to a 14.7 mm wide slit as a reproduction of their single-slit experiment. First of all we have investigated the question of chaotic behavior but find no supporting evidence for this. Thus unless the critical amplitude is surpassed we always find a causal behavior.  One seemingly solid result of their experiments was the existence of a limiting deflection angle that they suggest to be a generic feature caused by walker-wall interaction regardless of the specific experimental situation. While we do find that the  deflection never seems to become larger than about $\pm$ 70$^{\circ}$, our experiments show that this is by no means a dominating feature in the probability density function as seen in their experiments. On the contrary, as for the narrow slits we find much more complicated and fascinating structures. Finally we take a closer look at the back-bending tracks observed in their double-slit experiment. We have found this feature for a semi-broad  slit but also a related feature that arises both for the narrow single-slit and for the double-slit in a symmetric fashion for large deflection angles and is not related to the presence of any nearby structures as e.g. a second slit. 

Finally, no attempt is made to compare with existing theoretical models \cite{model1,model2,overview}, since our measurements show the system to be extremely sensitively dependent on so many parameters.

\section{Experimental details}
\begin{figure}
\hspace{-.15cm}\mbox{\epsfig{figure=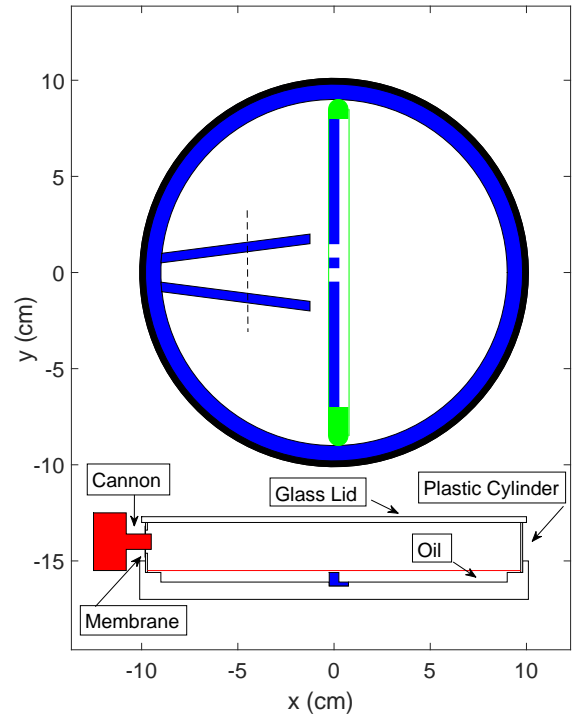,width=11cm}}
\caption{(Color online) Top: Top-view of the container, showing movable barrier, the accelerator and the elevated part of the bottom at the rim that  quenches rim generated waves. Bottom: Side-view, showing the plastic cylinder with the membrane and the cannon, the glass lid, and the oil surface. Dashed line shows the length of the shortened accelerator used in later experiments.}
\label{tray}
\end{figure}

Here we present the most important experimental features with a more detailed exposition given in App.~A. The innermost part of the container of 18 cm in diameter is surrounded by a 5 mm high and 7 mm broad shelf for depression of rim generated waves, making the system about twice as large as that of Couder and Fort. It is supplied with a built-in accelerator (also 5 mm high) and a central groove with a 5 mm broad double-slit barrier of height 5 mm that may be moved from outside (see Fig.~\ref{tray}). In later experiments the accelerator was shortened to end 37 mm before the barrier.  

The system is hermetically closed with a glass lid resting on a transparent plastic cylinder, although a plug allows for slow diffusive pressure equalization. A horizontal drops-on-command cannon \cite{Harris2} enters the side of the plastic cylinder through a hole sealed with a rubber membrane.  This configuration was chosen after it was observed that even tiny changes in the enclosed volume due to the vibration would result in a  build-up of air-flow disturbing the drop trajectories. An advantage of the setup is that it allows us to run for extended periods only opening the cell to change the barrier or change the fluid height.

The fluid used is  silicone oil of density $\rho$ = 950 kg m$^{-3}$, nominal viscosity $\nu$ = 20 $\times$ 10$^{-6}$ m$^{2}$ s$^{-1}$, and surface tension $\sigma$ = 0.0206 N m$^{-1}$, all at 25 $^{\circ}{\rm C}$. In accordance with this choice of fluid all measurements are performed at f = 80 Hz. The critical acceleration for the Faraday instability $\gamma_F$ is determined by looking for the spontaneous generation of surface waves. 

The detection of waves is made by observing reflected light from a circle of light emitting diodes (LED's) which in later experiments is supplemented with diodes placed in fixed positions giving reflections from close to the center but from outside the normal path of tracks. This latter addition allows us to monitor the status of each individual track. After temperature stabilization has been obtained, the critical acceleration $\gamma_F$ was determined  with an uncertainty of  $\pm$ 0.05 \% as was the driving acceleration $\gamma$. While the uncertainty given for $\gamma$ reflects our ability to keep $\gamma$ long-time constant, $\gamma_F$ is determined in the following fashion. First a rough measurement is performed. Then the Faraday waves are excited again and the system brought to a steady state just above $\gamma_F$. $\gamma$ is lowered in small steps until the waves die away and the whole process repeated a few times. This is a tedious procedure since the decay time $\tau \sim T_D(\gamma_F/(\gamma_F-\gamma))$, with T$_D \sim 1/\nu k_F^2$ and $k_F=2 \pi / \lambda_F$, is of the order of 1 min for the precision we aim at  \cite{Eddi}. However, the outcome is highly reproducible over time to within 0.2\%, with some small difference due to change of oil. Finally the measured wavelength of the Faraday waves is $\lambda_F \sim 4.8$ mm.

An important parameter characterizing the behavior of the freely moving droplet is the so-called memory factor proportional to $\Gamma^{-1} = \gamma_F/(\gamma_F-\gamma)$ that provides an estimate of how many past droplet impacts still have an influence on the present wave field pattern \cite{Eddi}. Since size effects and non-linear effects become important close to the critical acceleration, where the memory factor diverges, it is, however, unclear whether this factor gives a true and meaningful picture of droplet wave field behavior here. Discussion of the memory factor is therefore omitted in the rest of the paper (see also App. D).

\begin{figure}[ht]
\includegraphics[width=8.6cm]{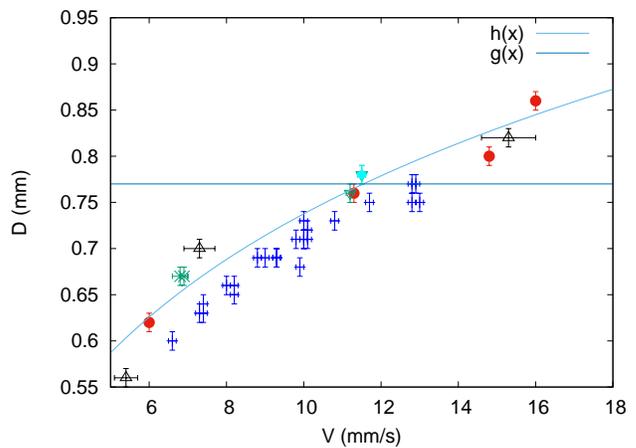}
\caption{(Color online) Plot of corresponding pairs of droplet size D and critical velocity v$_{imp}$ showing the investigated range of these properties. Our measurements ($\cdot$). Other measurements: Proti\`{e}re  {\it et al.} \cite{protiere} ($\triangle$), Mol\'{a}$\check{\rm c}$ek and Bush \cite{molacek} ($\bullet$), Pucci  {\it et al.} \cite{BushNew} ($\ast$),  Pucci  {\it et al.} \cite{Pucci} ($\triangledown$), Eddi  {it et al.} ($\diamond$) \cite{Eddi}, and h(x) fit thereto. The horizontal line g(x) corresponds to the upper limit for chaotically bouncing droplets.} 
\label{V-D}
\end{figure}

Tracks are recorded with a free-running (27.02 frames/s) digital video camera DMK 23UP031 set to spatial resolution 1500X1600 and analyzed by an adapted MATLAB tracking program \cite{tracking}. An estimate of the droplet size D can be obtained from the fitting routine showing a variation in a single run of less than $\pm$ 2\%, with a variation of at most $\pm$ 5\% in between runs. In contrast to Refs.~\cite{CouderFort,Andersen,BushNew} we use a new droplet for each track in order to have a fully automated system.  Reproducibility is therefore very important. While relative changes in droplet size are easily distinguished, the absolute scale factor is more difficult to ascertain. We therefore rely on the free speed close to $\gamma_F$ to characterize the droplets since this can be determined with high precision from the recordings. The phase-diagram for the droplet behavior is rather complicated (for an introduction see  e.g. \cite{WindWillassen}, \cite{Bush2015}, and \cite{molacek}) with the regime of a chaotic bouncing mode as the most interesting. As most of our measurements are in this regime, we document some features of the system performance here in App.~B. From the velocity measurements shown there, we derive the plot shown in Fig.~\ref{V-D} of corresponding pairs of droplet size and impact velocity v$_{imp}$ for $\gamma$ between 99\% and 100\% of $\gamma_F$. Both quantities are given as averages over 1 cm of tracks with each entry representing more than 100 droplets with the uncertainties representing the spread in data. The plot furthermore shows available data from other authors for comparison \cite{protiere,molacek,BushNew,Pucci,Eddi}. As most of these data are taken much further from the critical line where the droplets move noticeably slower, we have used the authors own extrapolation to the regime of interest in the comparison (footnote Ref.~\cite{unknownT}). A fit h(x) to these data is also shown (see footnote Ref.~\cite{FitCurve}) with our data lying in general about 4\% below. It is highly conceivable that this difference is due to the uncertainty in our measurement of the absolute scaling constant used in our droplet size determination.  
From the augmented phase-diagram, also shown in App. B, we can determine the maximum size of the droplet while still operating inside the chaotically bouncing regime. This is represented by the horizontal line g(x) in Fig.~\ref{V-D}. As seen most of our measurements are comfortably inside this regime even by a large margin, covering an extensive part of the regime while also overlapping the regime investigated by  Pucci  {\it et al.} \cite{BushNew}. Finally, since our uncertainties are always of the order shown here, we omit these in what follows below.

\noindent
\section{Experimental results}
We have mapped the critical forcing acceleration as a function of oil temperature T$_{oil}$ centered at an absolute temperature of 25.1  $\pm$ $0.1\,^{\circ}{\rm C}$. Since  $\gamma_F$ = $ 2^{4/3}(\rho/\sigma)^{1/3}\mu(2\pi f)^{5/3}$ is directly proportional to the viscosity \cite{Edwards}, the large temperature coefficient of this quantity \cite{rhodorsil} makes the dependency rather strong ($\sim$ 0.06g K$^{-1}$, see also Ref.~\cite{HarrisBush}). In order to keep the change in $\gamma_F$ within $\pm 0.1\%$ one has to keep the temperature from varying more than about $\pm$ 30 mK.

Since also  the excited wave amplitude and the memory time depend on the viscosity, it is not enough to rely on keeping the ratio $\gamma/\gamma_F$ constant. We have therefore chosen to enforce a tight control of T$_{oil}$, preferably keeping this within 5 mK in individual scans and inside a 15 mK window for all experiments reported here. Also we monitor the temperature  T$_L$ of the glass lid, keeping this from deviating more than about 500 mK from the oil temperature to keep possible air flow at the oil surface at a minimum \cite{Andersen3,BushNew}. Too large a difference in temperature is observed to give rise to curvature in the otherwise straight parts of the tracks on the outlet side (see App. C, Fig.~\ref{tempstab}). Air motion  seems in fact to be the main reason for the problems afflicting the early investigations. The tight temperature control moreover improves reproducibility considerably as the actual fluctuations in $\gamma_F$ become much smaller than the measuring accuracy.

\begin{figure}[ht]
\includegraphics[width=8.6cm]{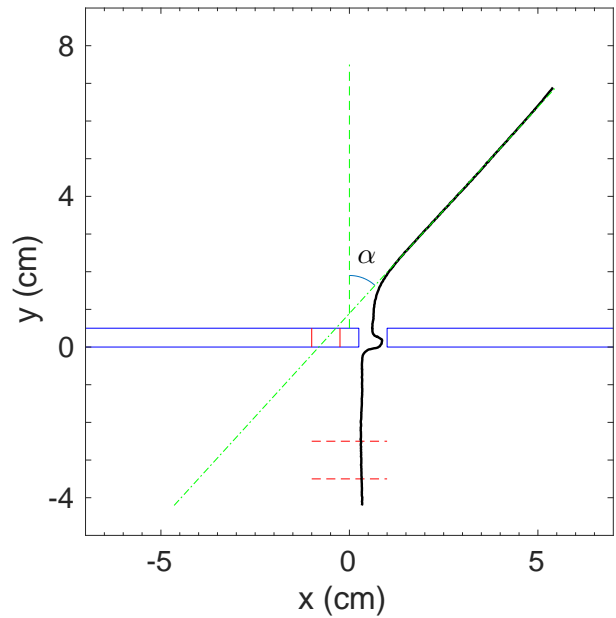}
\caption{(Color online) Representative track showing quality of fit to the straight part after passage and the definition of $\alpha$. The impact parameter x$_{imp}$ is defined as the average value of the x-coordinate between the lines at -3.5 and -2.5 cm and referred to the position of the center of the slit structure.}
\label{DefAlphaXimp}
\end{figure}

Two important quantities for the analysis are the deflection angle $\alpha$ and the impact parameter x$_{imp}$. These are defined in Fig.~\ref{DefAlphaXimp} which also shows a fit to the straight part of the track after passage.  The quantities $\alpha$ and x$_{imp}$ are determined within  $\pm 0.3^{\circ}$ and $\pm 20 \mu$m respectively, although the absolute zero for  x$_{imp}$ is only determined within $\pm 0.1$mm.

\subsection{Single-Slit}

We start by concentrating on an overview of single slit experiments. Four different slit widths have been investigated (5 mm, 7.3 mm, 7.5 mm, and 14.7mm). Here we present data obtained by  changing either the velocity (drop size), the driving acceleration, or the fluid height $h_1$ (measured with an uncertainty of $\pm$ = 0.03 mm) over the barrier, respectively, while keeping the other parameters constant. 

Before we go into a discussion of our results, we will describe our choice of presentation of data.  First, we present the geometric distribution of tracks that pass through the slit structure as seen in Fig.~\ref{5mmDepthVar}, left column. This leads to the correlation between  x$_{imp}$ and deflection angle $\alpha$, which in our case is completely sharp and reproducible (causal/deterministic) in all but the most extreme cases where the small but finite spread in drop-size becomes an issue,  Fig.~\ref{5mmDepthVar}, center column. Couder and Fort presented their data in a histogram  of the probability distribution of $\alpha$, but in most cases the number of tracks is far too small to make such a presentation meaningful. However, using the fact that the correlation is sharp we can get a more detailed representation of the distribution by fitting a continuous curve through the (x$_{imp}$,$\alpha$) correlation (see footnote Ref.~\cite{PDF}). This curve can then be projected onto the $\alpha$ axis to reveal more features of the distribution P($\alpha$),  Fig.~\ref{5mmDepthVar}, right column, than is possible directly with a histogram of the actual data points \cite{BushNew}. Such plots are presented in a few figures to stress the diversity of possible patterns, with histograms of the data distribution added in Fig.~\ref{7.5mmVelVar} and Fig.~\ref{S18og19a}. Note that some of the fine structures are obviously artifacts of the procedure and should be ignored. We should also point out that the procedure as detailed in  footnote Ref.~\cite{PDF} only produces a likely P($\alpha$) based on the data and the symmetry of the setup, not necessarily the correct P($\alpha$), and as such should be taken with a grain of salt.

\subsubsection{Dependence on fluid height}

\begin{figure*}[ht]
\includegraphics[width=17.2cm]{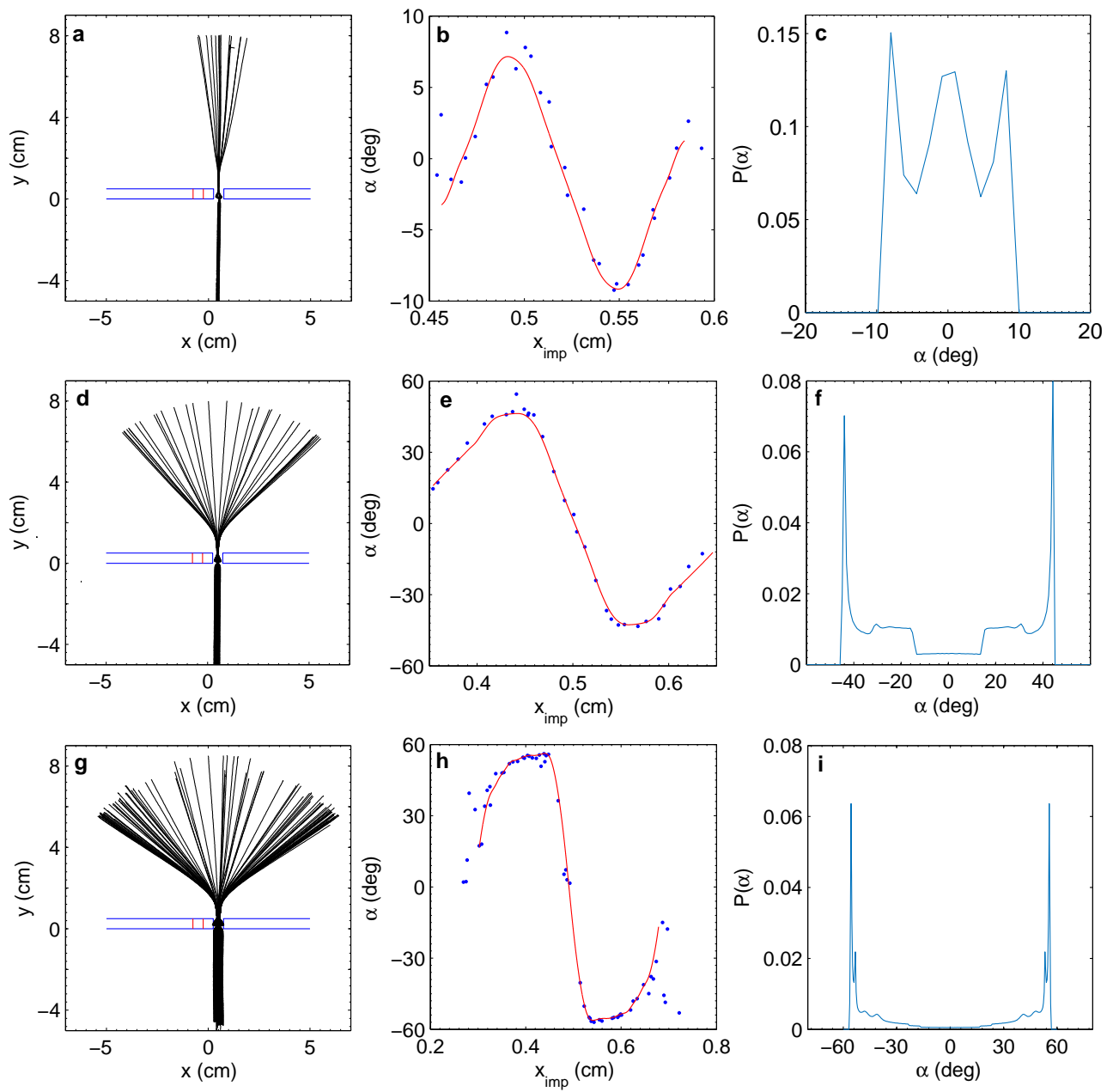}
\caption{(Color online) Track dependence on fluid height, 5 mm slit. From top to bottom: h$_1$= 0.51 mm, 31 tracks, v$_{imp}$ = 10.1 mm/s; h$_1$= 0.81 mm, 33 tracks, v$_{imp}$ = 10.6 mm/s; h$_1$= 0.96 mm,  108 tracks, v$_{imp}$ = 9.9 mm/s. For all the drive amplitude  is $\gamma/\gamma_F = 0.999$. Left column: tracks for a range of  x$_{imp}$; middle column:  ($\cdots$) $\alpha$ versus x$_{imp}$ and ($-$) fit; right column: P($\alpha$) versus $\alpha$.}
\label{5mmDepthVar}
\end{figure*}

We now discuss the influence of varying the fluid height h$_1$ over the barrier for a 5 mm slit for droplets of size 0.67 mm having  velocities of 10 mm/s with results displayed in Fig.~\ref{5mmDepthVar}. 
The impact parameter x$_{imp}$ is determined from the video files with sub-pixel precision from markers on the bottom of the container and on the barrier by a MATLAB program. Note that we measure x$_{imp}$ from the center of the barrier structure.

In the left column of Fig.~\ref{5mmDepthVar} is displayed the development of the tracks as the impact parameter x$_{imp}$ is swept across the slit with normalized driving acceleration $\gamma/\gamma_F = 0.999$. From top to bottom the fluid height  h$_1$ over the barrier has the values of  0.51, 0.81 mm, 0.96 mm respectively.  In the middle column the corresponding deflection angles  $\alpha$ are shown as a function of  x$_{imp}$. In all cases the relation between $\alpha$ and x$_{imp}$ approximates a whole cycle of a somewhat distorted sinusoidal oscillation but, evidently, is completely causal in contrast to the statement by Couder and Fort \cite{CouderFort} that there is no correlation. 

It is furthermore clear that both the range of possible impact parameters x$_{imp}$ for passage of the slit and the fan of corresponding deflection angles $\alpha$ are strong functions of the fluid height over the barrier. At the lowest height h$_1 = 0.51$ mm (Fig.~\ref{5mmDepthVar}a) droplets, moving slower than chosen here, are quite simply not making it through the narrow slit as the window of possible values of x$_{imp}$ narrows down. 

At the largest h$_1 = 0.96$ mm (Fig.~\ref{5mmDepthVar}g and h) we also see a new phenomenon at the outskirts of the possible impact parameters for passages. In a small range gaps are observed where droplets are reflected, interspersed with successful passages having widely diverging deflection angles though still keeping within the maximal deflection set by the continuum of x$_{imp}$ values. Remarkably enough these tracks are still completely reproducible. Also displayed in the middle column are curves from symmetrized fits to the data. The rightmost column shows the density distribution functions P($\alpha$) corresponding to the fitted curves. Already here it is clear that even though at the lowest value of  h$_1 = 0.51$ some resemblance is seen to the results by Couder and Fort  \cite{CouderFort} with P($\alpha$) displaying a central peak and two side peaks, quite different forms are also possible with the loss of the central peak at higher fluid levels.
 
\subsubsection{Dependence on droplet velocity for 5 mm slit}

Next we turn to the influence of varying the droplet velocity v$_{imp}$, or in this case equivalently the droplet size D, while keeping the fluid height and driving acceleration constant. 
\begin{figure}[th]
\includegraphics[width=8.6cm]{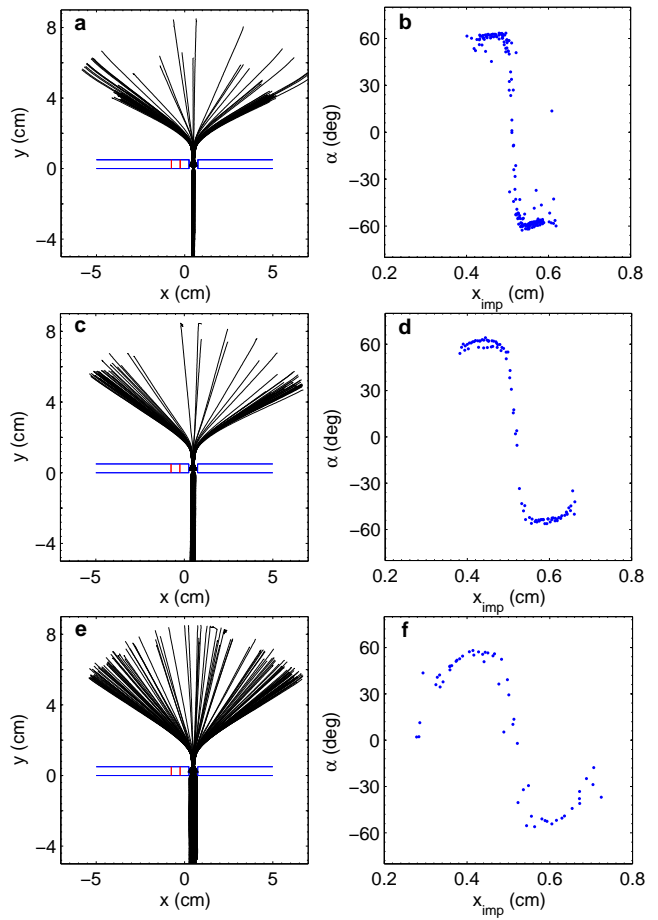}
\caption{(Color online) Track dependence on droplet velocity for 5 mm slit. From top to bottom: 162 tracks, v$_{imp}$ = 8.2 mm/s;   92 tracks, v$_{imp}$ = 8.8 mm/s; 133 tracks, v$_{imp}$ = 9.8 mm/s. Drive amplitude  $\gamma/\gamma_F = 0.9995 \pm 0.0008$, h$_1$= 0.96 mm. Left column: tracks for a range of x$_{imp}$; right column: $\alpha$ versus x$_{imp}$. Please note that these figures in all cases represent more than one sweep across the slit.}
\label{5mmVelVar}
\end{figure}  
The left column in  Fig.~\ref{5mmVelVar} shows the collection of tracks as the impact parameter x$_{imp}$ is swept across the slit for fixed values of  h$_1 = 0.96$mm and drive amplitude $\gamma/\gamma_F = 0.9995$, while the right column shows the corresponding deflection angles  $\alpha$ as a function of  x$_{imp}$. From top to bottom the velocities are respectively 8.2, 8.8, and 9.8 mm/s. (The shorter tracks are due to the total number of frames being set too low.)
At the lowest value of v$_{imp} =$ 8.2 mm/s  the range of possible impact parameters x$_{imp}$ is rather narrow (approximately 2 mm) with a maximal deflection of close to 60$\,^{\circ}$ and a very rapid change in deflection over a range of approximately 100 $\mu$m in impact parameter. As the velocity goes up, the range of impact parameters increases but the fan of angular distribution decreases slightly in width at the same time, while the overall features soften up. It is worth noting here that we again find a lower limit on the droplet velocity that allows for traversing the slit, with this limit being a decreasing function of the fluid height over the barrier. The corresponding  P($\alpha$) will of course all look like that at the bottom of Fig.~\ref{5mmDepthVar} with no central peak and peaks at the values of $\alpha$ at maximal deflection.
 
\begin{figure}[th]
\includegraphics[width=8.6cm]{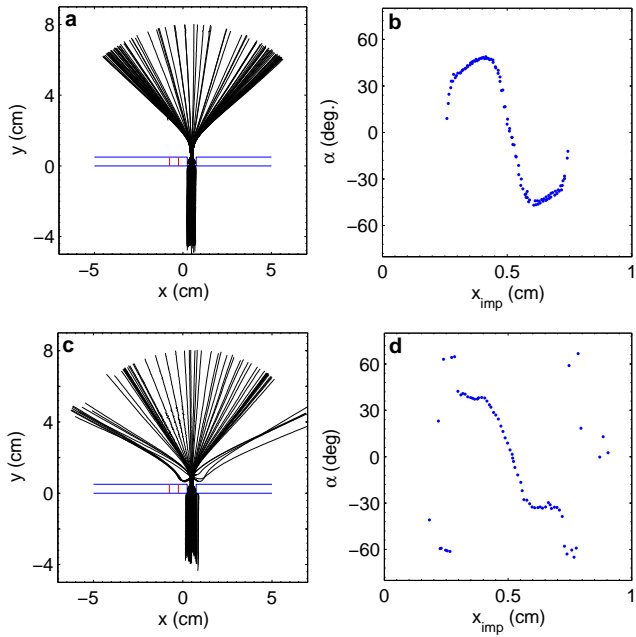}
\caption{(Color online) Track dependence on droplet velocity, 5 mm slit. From top to bottom: 103 tracks, v$_{imp}$ = 11.4 mm/s;   67 tracks, v$_{imp}$ = 13.0  mm/s. Drive amplitude  $\gamma/\gamma_F =0.999$,  h$_1$= 0.81 mm.  Left column: tracks for a range of x$_{imp}$; right column: $\alpha$ versus x$_{imp}$.}
\label{5mmVel2Var}
\end{figure}

In Fig.~\ref{5mmVel2Var} tracks, obtained at a slightly lower value h$_1$ = 0.81 mm over the barrier, for velocities of 11.4 mm/s and 12.4 mm/s are displayed. 
These tracks follow the trends of the previous figure. Notice that at these high values of the droplet velocity, the change in  h$_1$ plays a minor role. At the highest velocity we again see the sporadic passing of droplets at the ends of the range in impact values. However, some of these tracks show a strange behavior in a sudden back-tracking toward the barrier just after passage before veering off in straight tracks which are mirror images (y-axis symmetry) of the bulk of nearby tracks. Interestingly enough these tracks are also reproducible. While P($\alpha$) for the velocity 11.4  mm/s looks like the preceding cases, the sporadic tracks seen at  the velocity 12.4 mm/s if included would cause weaker satellite peaks to occur in P($\alpha$) at around $\pm 60\,^{\circ}$ outside the peaks due to the maximum value of $|\alpha| \approx 40\,^{\circ}$ from the continuum range of $x_{imp}$).

From a consideration of Figs.~\ref{5mmDepthVar}, \ref{5mmVelVar}, and \ref{5mmVel2Var} it is furthermore obvious that the effective slit opening depends not only on the Faraday wavelength but also is strongly dependent on the fluid height h$_1$ and the droplet velocity. In our presentation we have therefore refrained from  normalizing the impact parameter with the slit width to avoid giving a false impression that simple data-collapse is possible. The dependence on  h$_1$ is the natural consequence of the lack of sharp boundary conditions for the droplet associated wave pattern  as the damping by the barrier structure is weakened with rising fluid level. The dependence on the droplet velocity on the other hand is more akin to an energy barrier effect. 

\subsubsection{Velocity dependence for 7.5 mm slit}
\begin{figure*}[th]
\includegraphics[width=17.2cm]{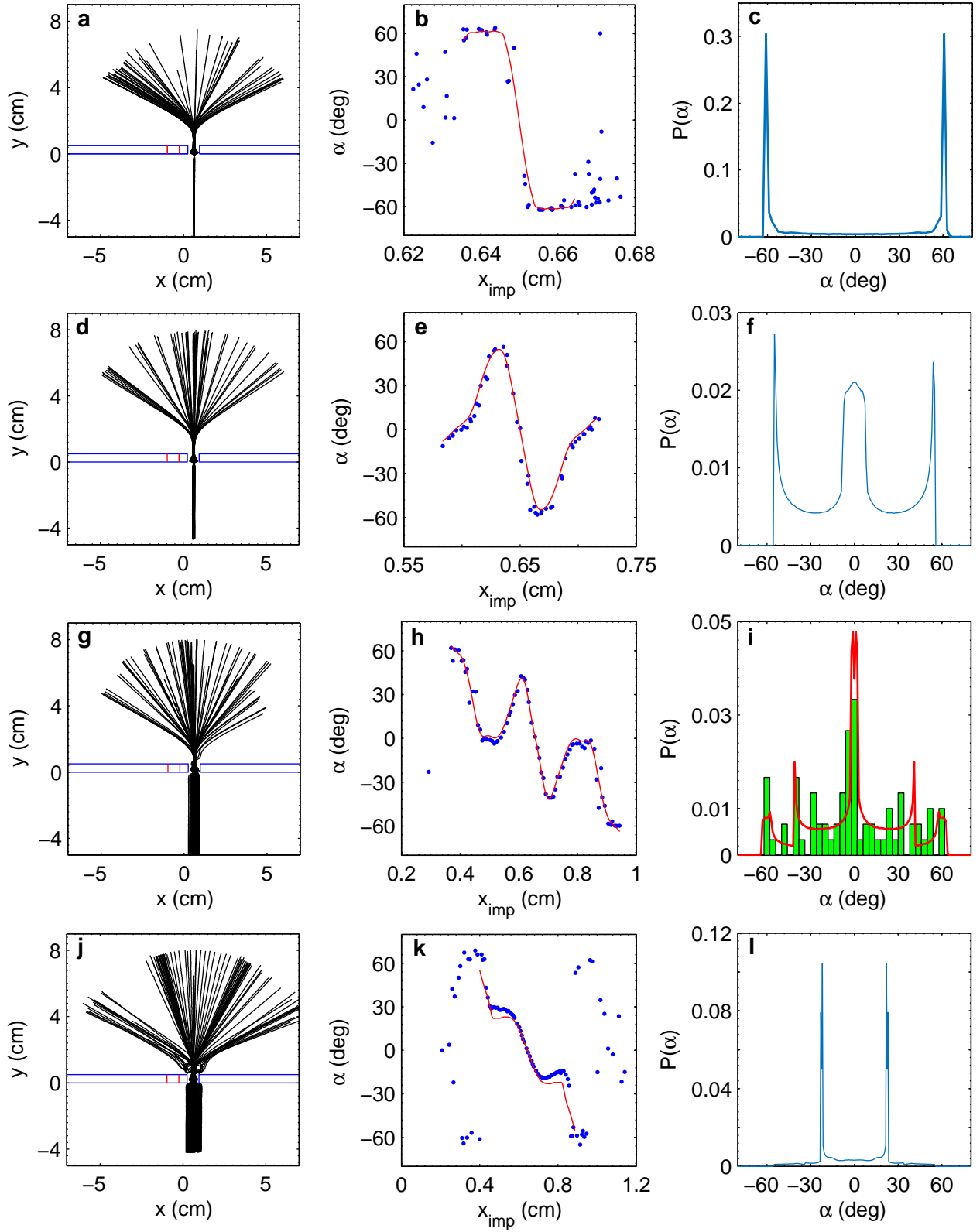}
\caption{(Color online) Track dependence on droplet velocity for 7.5 mm slit. From top to bottom: 60 tracks, v$_{imp}$ = 7.4 mm/s,  $\gamma/\gamma_F = 0.999$;   51 tracks, v$_{imp}$ = 8.0 mm/s; 73 tracks, v$_{imp}$ = 10.0 mm/s; 94 tracks, v$_{imp}$ = 12.8 mm/s, $\gamma/\gamma_F = 0.995$. h$_1$= 0.62 mm. Left column: tracks for a range of x$_{imp}$; middle column: ($\cdots$) $\alpha$ versus x$_{imp}$ and ($-$) fit; right column: P($\alpha$) versus $\alpha$ including in (i) a normalized histogram of the data-set.}
\label{7.5mmVelVar}
\end{figure*}

The velocity dependence has also been investigated for a 7.5 mm slit with some results presented here in Fig.~\ref{7.5mmVelVar}.
From top to bottom the velocities are  7.4 mm/s with $\gamma/\gamma_F = 0.999$,  and 8.0 mm/s, 10.0 mm/s, and 12.84 mm/s, all with $\gamma/\gamma_F = 0.995$. The value of  h$_1$ = 0.62 mm in all cases. 

Comparing to the results for the 5 mm slit there are some similarities but also some striking differences. From the left column it is clear that the maximum value of $|\alpha|$ is nearly independent of the velocity following the trend seen before. Looking at the middle column,  which shows the dependence of $\alpha$ on the impact parameter $x_{imp}$, is more revealing. At the lowest velocity, the same narrow apparent slit width dominates the response and we again get the rapid flip of $\alpha$ within approximately 100 $\mu$m. At the highest velocity the response again becomes remarkably similar to that at the highest velocity for the 5 mm slit. In between we see that the pattern has developed from the simple one-cycle pattern seen for the 5 mm slit to a more complicated pattern including more cycles. 

The right column shows that the P($\alpha$)'s derived from the fits displayed in the middle column go from a simple pattern with no center peak, over a pattern with a central peak and a single pair of side peaks stemming from the maximum value of $|\alpha|$, then a central peak with two sets of side-peaks, while finally the central peak disappears and only the side peaks remain. In Fig.~\ref{7.5mmVelVar}i we have included a normalized histogram of the data-points to visualize the correspondence.

At the highest velocity we again encounter the phenomenon of sporadic traces including those where the droplet turns back in its track towards the barrier before veering off and resuming along a straight track at an angle of opposite sign from those of the nearby continuum. These sporadically occurring tracks have been neglected in the fitting  procedure due to their rapid sign change of deflection angle and intermittent nature.

Two immediate conclusions can be drawn from these measurements. Firstly, since the effective slit width shows up to be parameter dependent with the angular response also dependent on velocity, the appearance of more cycles is not due to a simple geometric wave number selection effect related to the width of the slit. The picture emerging is that of a much more complicated interplay between the wave field and the boundary conditions. Secondly some choice of parameters might accidentally result in a probability density function P($\alpha$) closely reminiscent of the single-slit quantum result.

\subsubsection{Velocity dependence for 14.7 mm slit}
\begin{figure}[th]
\includegraphics[width=8.4cm]{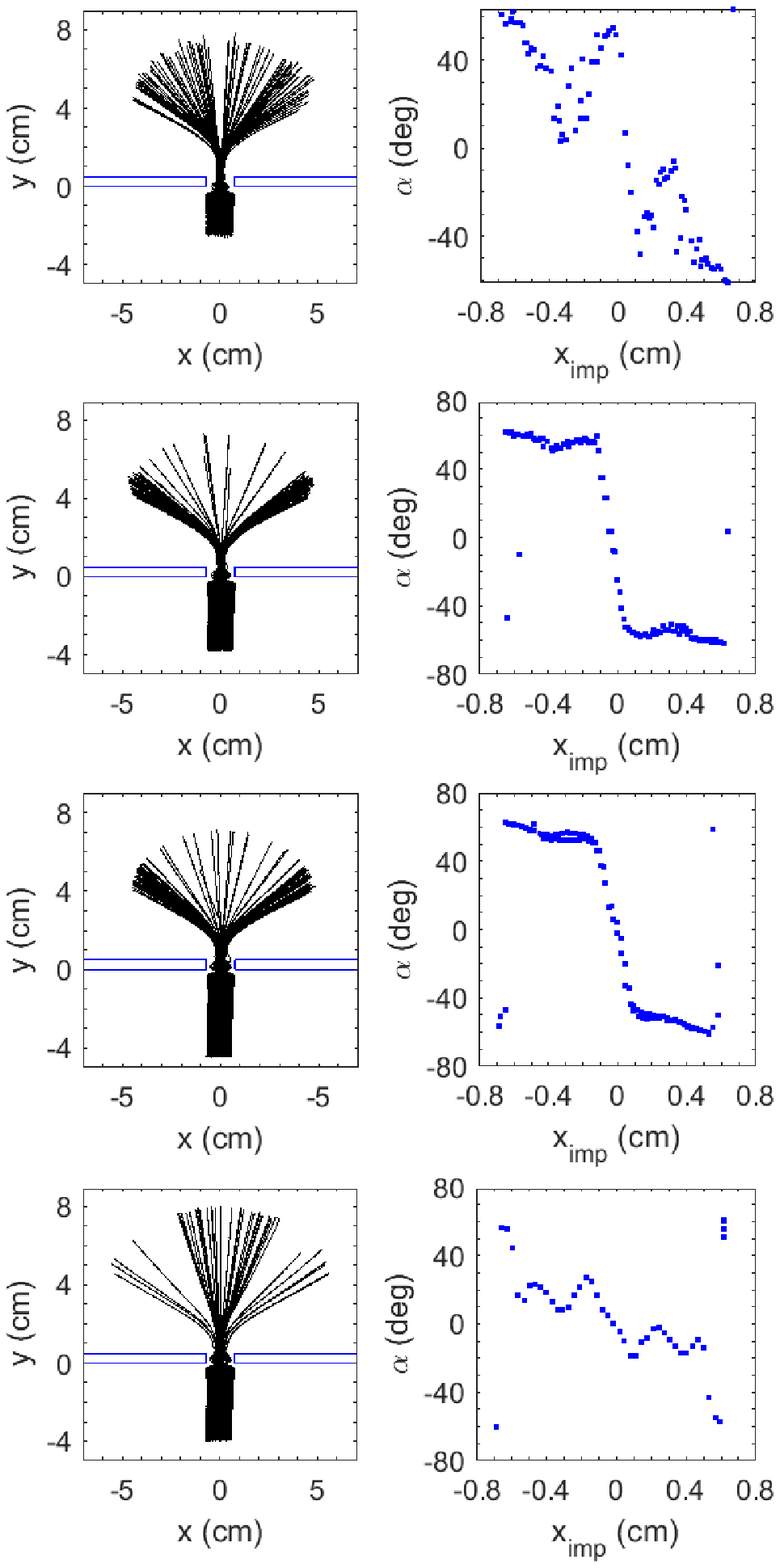}
\caption{(Color online) Track dependence on droplet velocity for 14.7 mm slit. From top to bottom: 62 tracks, v$_{imp}$ = 6.3 mm/s; 107 tracks, v$_{imp}$ = 6.8 mm/s; 102 tracks, v$_{imp}$ = 7.6 mm/s;  45 tracks, v$_{imp}$ = 8.7 mm/s. $\gamma/\gamma_F = 0.985$. h$_1$= 0.59 mm. Left column: tracks for a range of x$_{imp}$; right column: $\alpha$ versus x$_{imp}$.}
\label{14.7mmVelVar}
\end{figure}

Finally, the velocity dependence has been investigated for a 14.7 mm slit with some results presented in Fig.~\ref{14.7mmVelVar}.
From top to bottom the velocities are 6.3 mm/s, 6.8 mm/s, 7.6 mm/s, and 8.7 mm/s, all with $\gamma/\gamma_F = 0.985$. The value of  h$_1$ is 0.59 mm in all cases. Much to our surprise the outcome shows a much richer scenario than earlier observed. In fact the overall picture we get is very much like that for the narrower 7.3 and 7.5 mm slits again with a strong dependence on velocity. However, the droplet with the velocity 6.8 mm/s does behave very much like what was found by Pucci {\it et al.} \cite{BushNew} (their Fig.3b) for the same velocity.

\subsubsection{Dependence on acceleration}
\begin{figure*}
\includegraphics[width=17.2cm]{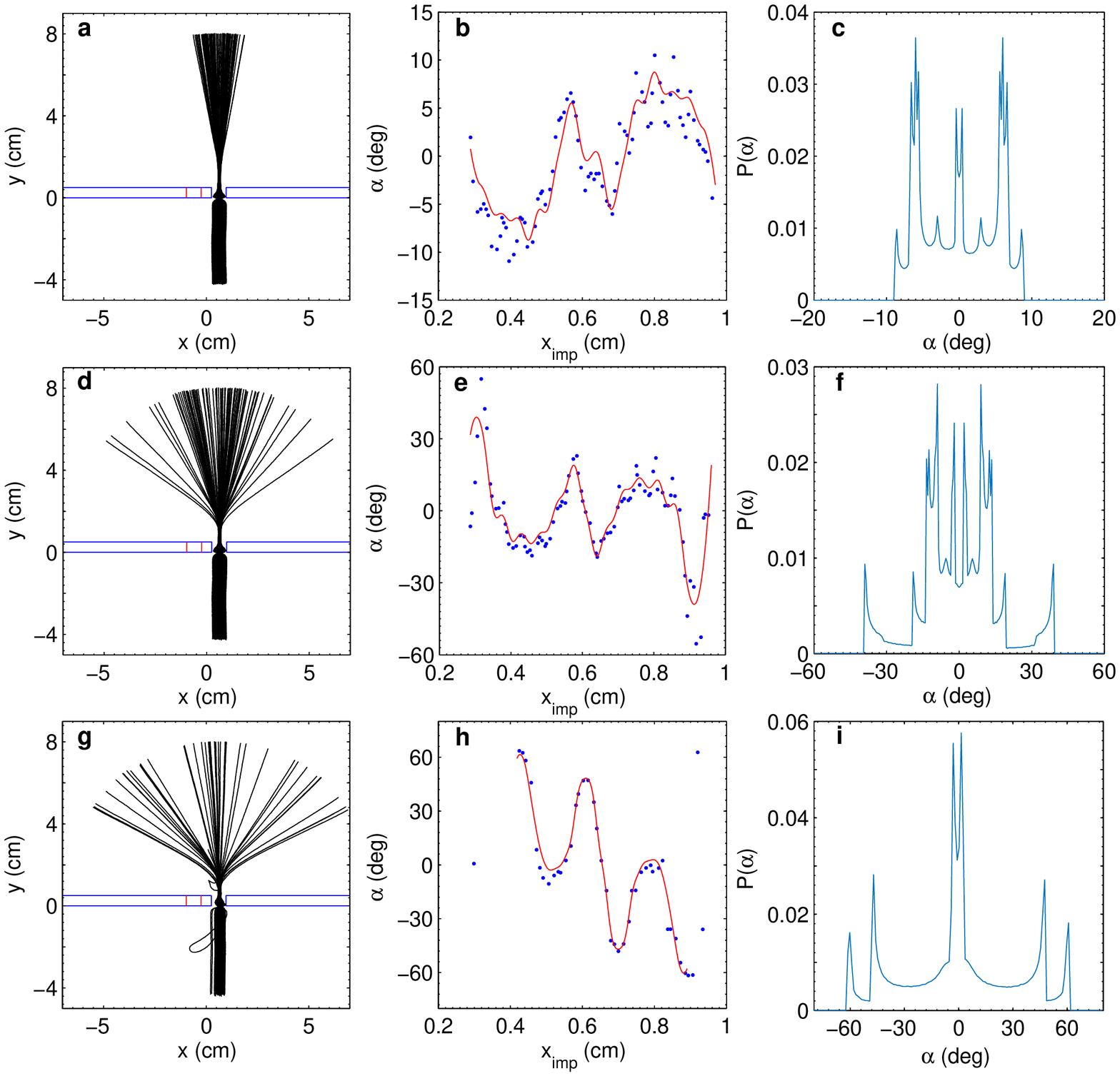}
\caption{(Color online) Dependence on acceleration, 7.5 mm slit. From top to bottom: 82 tracks, D = 0.69 mm, v$_{imp}$ = 9.8 mm/s, $\gamma/\gamma_F = 0.97$;   84 tracks, D = 0.71 mm, v$_{imp}$ = 9.9 mm/s,  $\gamma/\gamma_F = 0.99$;   44 tracks, D = 0.68 mm, v$_{imp}$ = 9.9 mm/s, $\gamma/\gamma_F = 0.999$. h$_1$= 0.61 mm. Left column: tracks for a range of x$_{imp}$;  middle column: ($\cdots$) $\alpha$ versus x$_{imp}$ and ($-$) fit; right column: P($\alpha$) versus $\alpha$.}
\label{7.5mmAmplVar}
\end{figure*}

To illustrate the change with drive amplitude we show in Fig.~\ref{7.5mmAmplVar}  some results obtained with the 7.5 mm slit. Here the droplet velocity is kept close to 10 mm/s while the fluid height  h$_1$ = 0.61 mm. 
Looking at the left column we see that the initially narrow fan of output tracks at  $\gamma/\gamma_F = 0.97$ opens up as the amplitude increases towards the critical value. The middle column shows that there is a definite shift in the angular symmetry, and the range of possible impact parameters becomes slightly restricted as the amplitude is increased. The latter effect is even more pronounced for the 5 mm slit. Furthermore,  the number of cycles in the angular response has increased. Thus the angular response as also evident from Fig.~\ref{7.5mmVelVar}h is drastically changed from what was seen at the more narrow 5 mm slit due to the increased space for the droplet associated wave pattern.

\begin{figure}
\includegraphics[width=8.2cm]{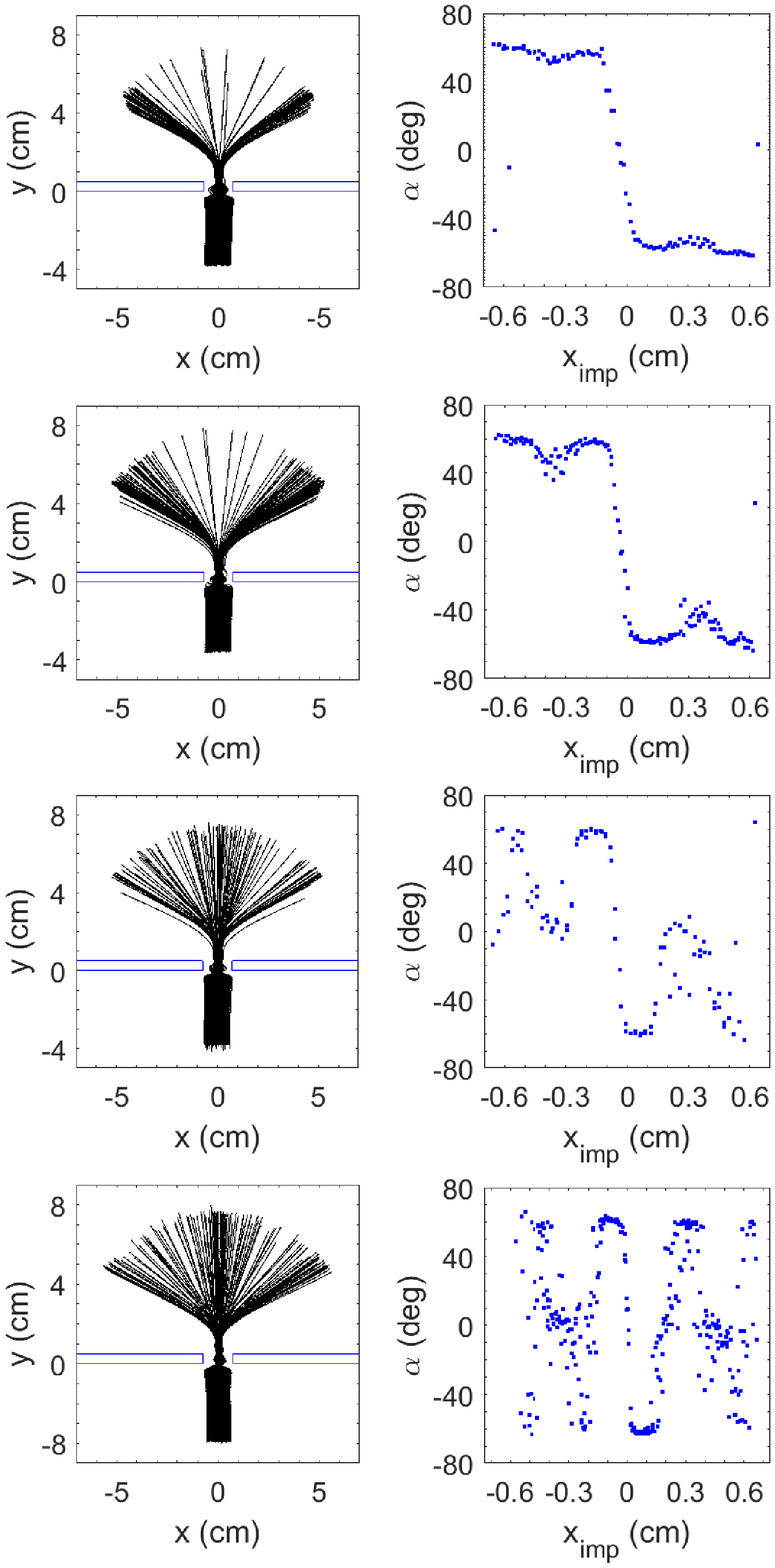}
\caption{(Color online) Track dependence on acceleration, 14.7 mm slit. From top to bottom: 107 tracks, v$_{imp}$ = 6.8 mm/s, $\gamma/\gamma_F = 0.985$;   156 tracks, v$_{imp}$ = 7.0 mm/s,  $\gamma/\gamma_F = 0.991$;   102 tracks, v$_{imp}$ = 7.1 mm/s, $\gamma/\gamma_F = 0.995$; 281 (135 shown) tracks, v$_{imp}$ = 7.3 mm/s,  $\gamma/\gamma_F = 0.998$. h$_1$= 0.59 mm. Left column: tracks for a range of x$_{imp}$;  right column: $\alpha$ versus x$_{imp}$.}
\label{14.7mmAmplVar}
\end{figure}

We have repeated these measurements for the 14.7 mm slit with results shown in Fig.~\ref{14.7mmAmplVar}.
As we increase the normalized drive from 0.985, we see two peaks centered around x$_{imp}$ $\pm$ 3 mm starting to develop in the comparatively flat portions of the plots of $\alpha$ versus x$_{imp}$. At 0.998 these peaks have reached all the way across. Thus we find that the pattern of more cycles appearing in the angular response seen for the 7.5 mm slit is repeated for the 14.7 mm slit.

\begin{figure}[th]
\includegraphics[width=8.6cm]{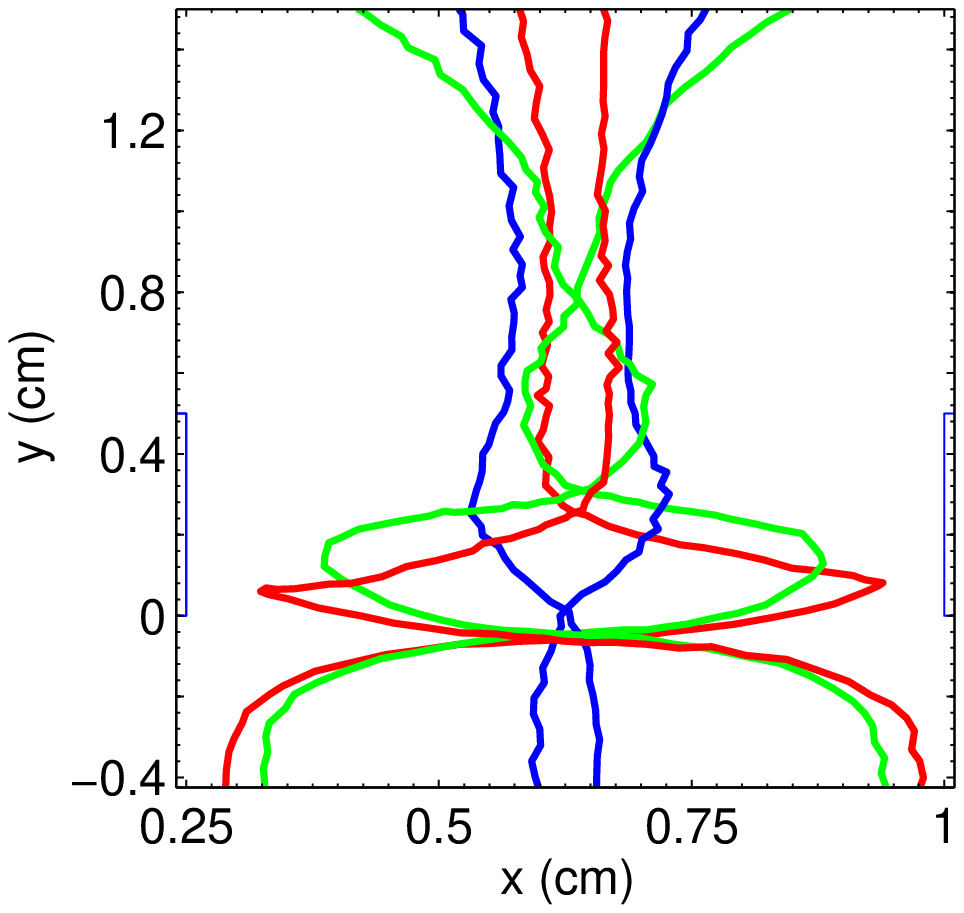}
\caption{(Color online) Magnification of the central area of Fig.~\ref{7.5mmAmplVar}d showing the complexity of the tracks inside the slit region. Note that the y-axis covers 2.5 times that of the x-axis.}
\label{ComplexPic}
\end{figure}

It is now clear that the simple picture, that tracks with impact parameters to the right of the center of the slit end up turning left and vice versa, given by Pucci {\it et al.} for a 14.7 mm broad slit in their Fig. 3, breaks down both for changes in  velocity, for higher accelerations, as well as for more narrow slits. This is illustrated in Fig.~\ref{ComplexPic} where the complex shifts in patterns inside the 7.5 mm slit can be seen as an example.
Three pairs of tracks have been chosen to show the symmetries encountered in the passage. The innermost tracks have one crossing of the center-line. The outermost tracks have two, and the tracks of the pair in between have three crossings each.

\noindent
\subsection{Double-Slit}
In this section we shall present results obtained in double-slit experiments where  a central piece of barrier of width 4.7mm is surrounded by a pair of either narrow (5mm) or broader (7.3mm) slits.

\begin{figure}
\includegraphics[width=8.6cm]{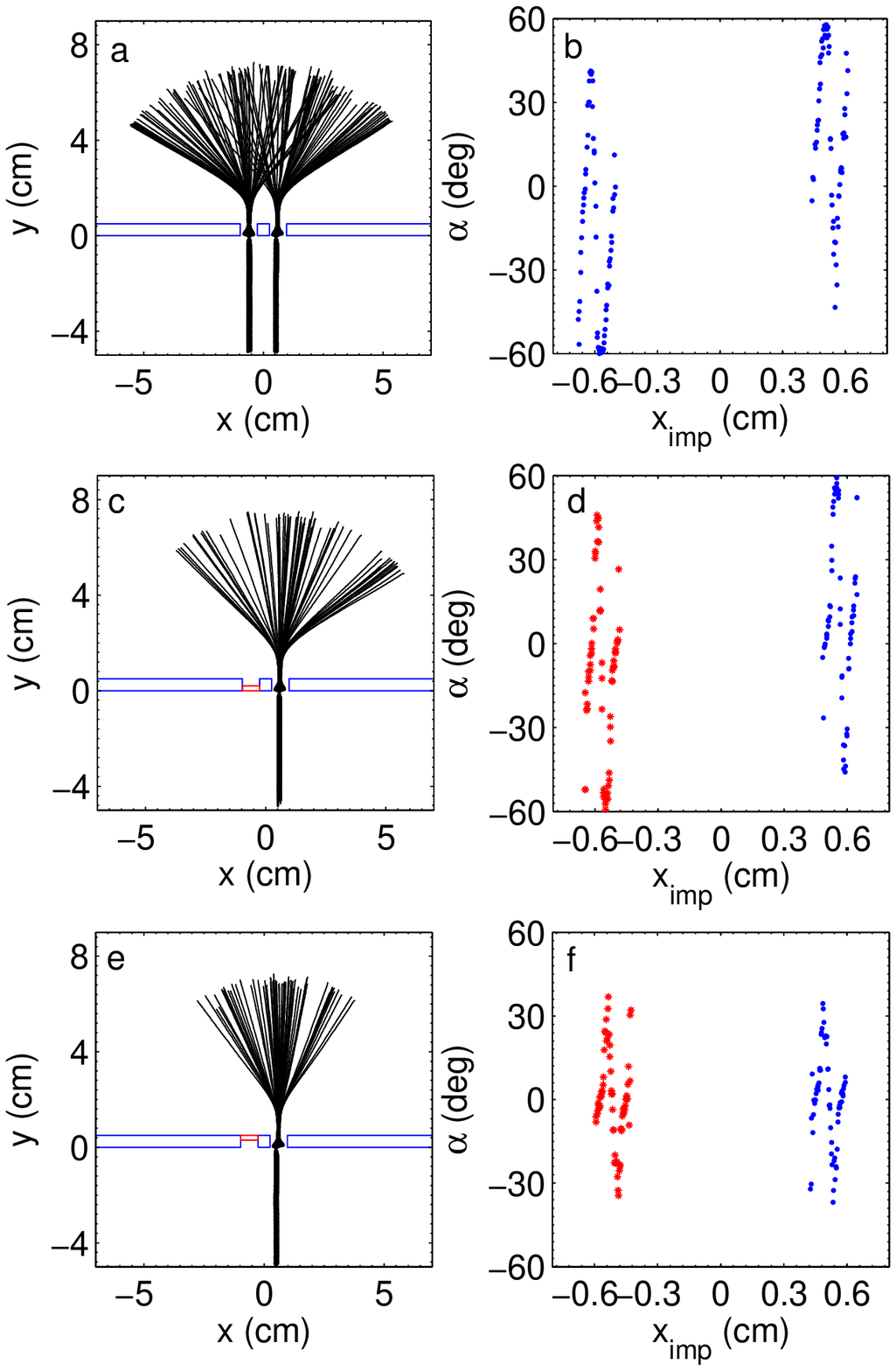}
\caption{(Color online) Effect of closing one slit, slit size 7.3 mm. Upper row: Double-slit experiment: (a) 129 tracks. (b) Corresponding $\alpha$ versus  x$_{imp}$ showing point-symmetry around the origin. Middle row: Left slit blocked flush at entrance with 2 mm plug. (c) 62 tracks. (d) ($\cdots$) corresponding $\alpha$ versus  x$_{imp}$. ($\ast$) Point-symmetric copy of the data showing the  same symmetry break as for the double-slit. Lower row: Left slit blocked flush at outlet with 2 mm plug. (e) 63 tracks. (f) ($\cdots$) corresponding $\alpha$ versus  x$_{imp}$. ($\ast$) Point-symmetric copy of the data showing no symmetry break as expected for a single-slit. $\gamma/\gamma_F = 0.996$, v$_{imp}$ = 8.3 mm/s, h$_1$= 0.71 mm.}
\label{Ang_dist}
\end{figure}

In Fig.~\ref{Ang_dist} upper row we show in (a) a plot of the tracks obtained by  sweeping x$_{imp}$ across both slits  from an experiment with slit openings 7.3mm, with only tracks from droplets that pass through the slits shown. Tracks are recorded for $\gamma/\gamma_F = 0.996$ and v$_{imp}$ = 8.3 mm/s. In (b) we show the corresponding deflection angles  $\alpha$ as a function of  x$_{imp}$. On one hand, we again find a strict causal behavior. On the other hand, the patterns are now skewed left and right for the left and right slit respectively, and thus only have point symmetry. This  indicates that some interference effect is indeed present as claimed by Couder et Fort \cite{CouderFort} and  also noted by Andersen {\it et al.} \cite{Andersen}, and Pucci {\it et al.} \cite{BushNew}.
                                              
To find the cause of this interference we have conducted double-slit experiments with slits opened or closed in various ways. In Fig.~\ref{Ang_dist} middle row we show the tracks obtained  and the resulting deflection angle $\alpha$ versus impact parameter x$_{imp}$ with a 2 mm deep plug blocking the input side of the barrier but leaving a substantial hole in the barrier seen from the far side.

The plot  Fig.~\ref{Ang_dist}d is seen to be very similar to that for the double-slit. To make the comparison easier we have included a point-symmetric copy ($\ast$). In contrast the corresponding plot  Fig.~\ref{Ang_dist}f for blocking flush with the output side shows a completely symmetric behavior as expected for a single-slit. Again we include a point-symmetric copy for comparison. 

In a similar experiment we take a look at faster drops, v$_{imp}$ = 11.1 mm/s, having correspondingly lower transit times, to test whether this time plays any role in the outcome.
\begin{figure}
\includegraphics[width=8.6cm]{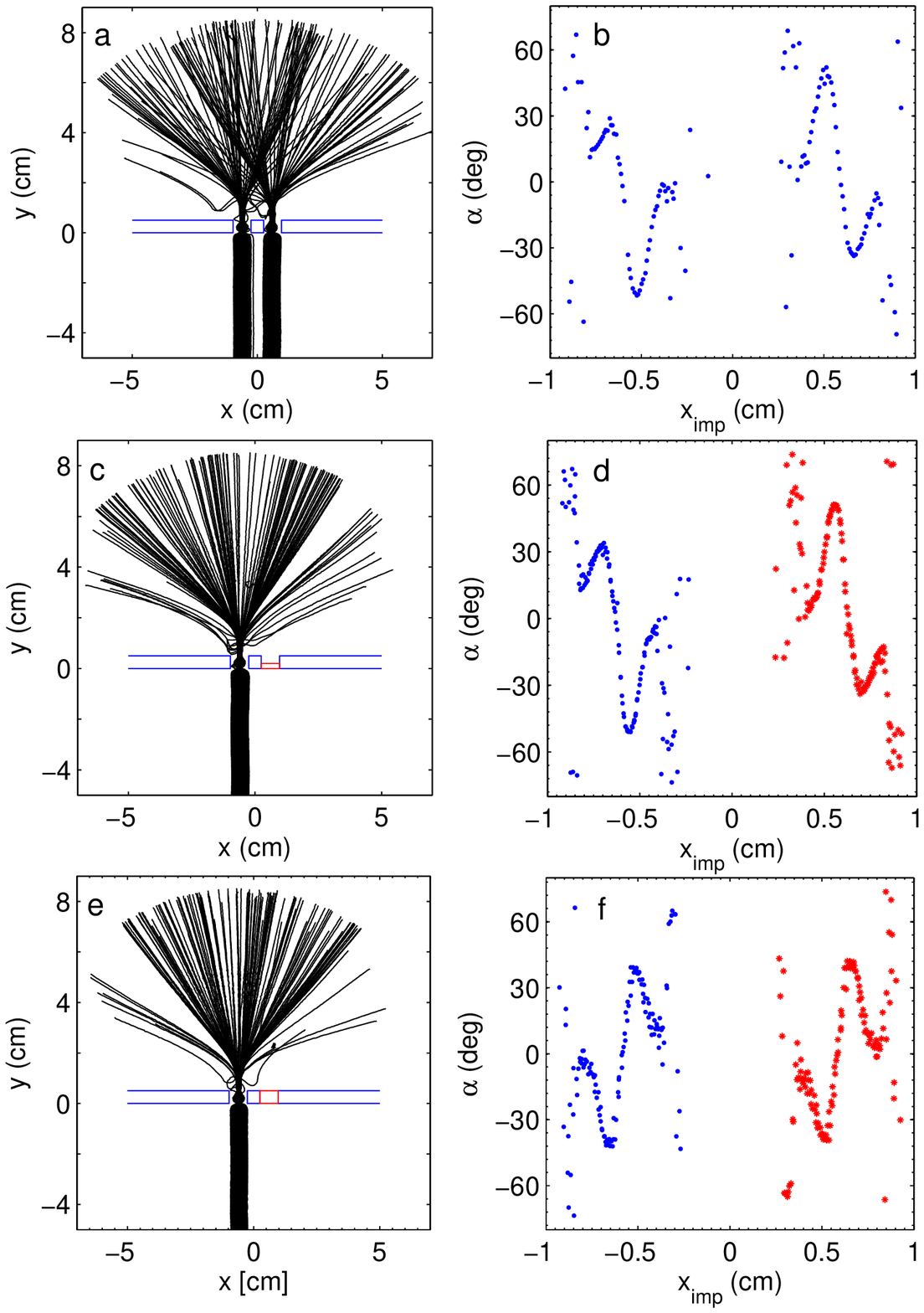}
\caption{(Color online) Effect of closing one slit, slit size 7.3 mm. Upper row: Double-slit experiment: (a) 132 tracks, v$_{imp}$ = 11.3 mm/s. (b) Corresponding $\alpha$ versus  x$_{imp}$ showing point-symmetry around the origin. Middle row: Right slit blocked at entrance with 2 mm plug. (c) 130 tracks, v$_{imp}$ = 11.2 mm/s. (d) ($\cdots$) corresponding $\alpha$ versus  x$_{imp}$. ($\ast$) Point-symmetric copy of the data showing the  same symmetry break as for the double-slit. Lower row: Right slit blocked flush at outlet with 5 mm plug. (e) 128 tracks, v$_{imp}$ = 11.0 mm/s. (f) ($\cdots$) corresponding $\alpha$ versus  x$_{imp}$. ($\ast$) Point-symmetric copy of the data showing no symmetry break as expected for a single-slit. $\gamma/\gamma_F = 0.999$,  h$_1$= 0.78 mm.}
\label{Ang_dist2}
\end{figure}
As seen from  Fig.~\ref{Ang_dist2} we reach the same conclusion as above.

A similar picture is obtained using the 5 mm slits. Here we have made measurements with both slits open, with a 2.5mm plug blocking the inlet of the right slit, or with the 2.5 mm plug blocking the outlet in either the left or the right slit.

\begin{figure}
\includegraphics[width=8.6cm]{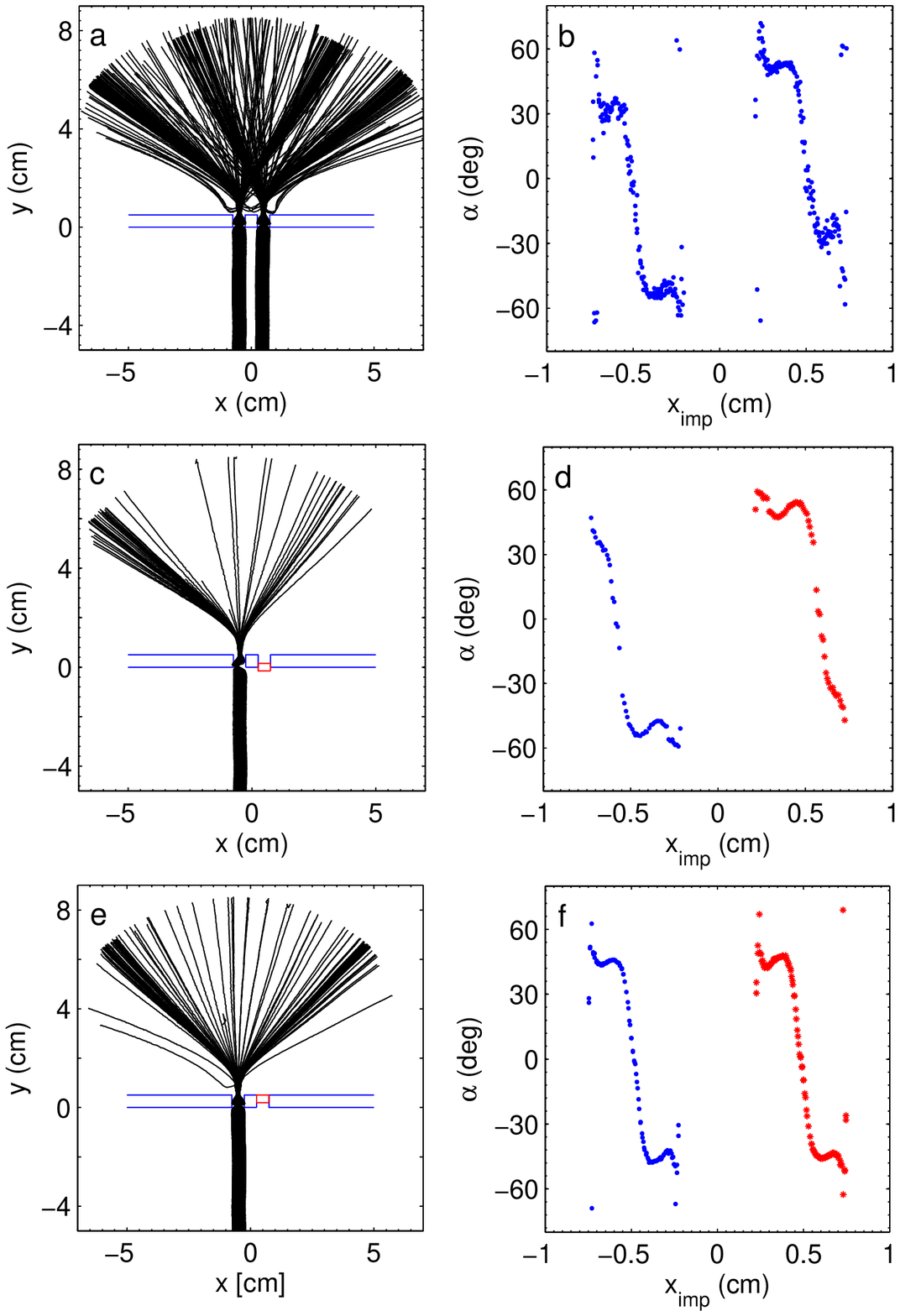}
\caption{(Color online) Effect of closing one slit. Slit size 5 mm. (a) 336 tracks through double-slit; (b) corresponding $\alpha$ versus x$_{imp}$. (c) Right slit blocked at entrance to slit, 55 tracks; (d) ($\cdots$) corresponding $\alpha$ versus x$_{imp}$ together with point-symmetric copy ($\ast$) showing the same symmetry break as for the double-slit experiment. (e) Right slit blocked flush with outlet side, 113 tracks; (f)  corresponding $\alpha$ versus x$_{imp}$ ($\cdots$) together with  point-symmetry copy ($\ast$). No symmetry break as expected for a single-slit. v$_{imp}$ = 11.1 mm/s, $\gamma/\gamma_F = 0.999$,  h$_1$= 0.71 mm.}
\label{Block}
\end{figure}

In Fig.~\ref{Block} we show the resulting tracks and the corresponding plots of angular deflection versus impact parameter for the different situations. Again it is clear that plugging the entrance does not influence the pattern of tracks through the other slit but plugging the outlet immediately brings us back to the single-slit situation. 

To illuminate further the cause of this interference effect let us take a closer look at the parts of the tracks that hit the middle block of the barrier.  

\begin{figure}
\centering
\mbox{\epsfig{figure=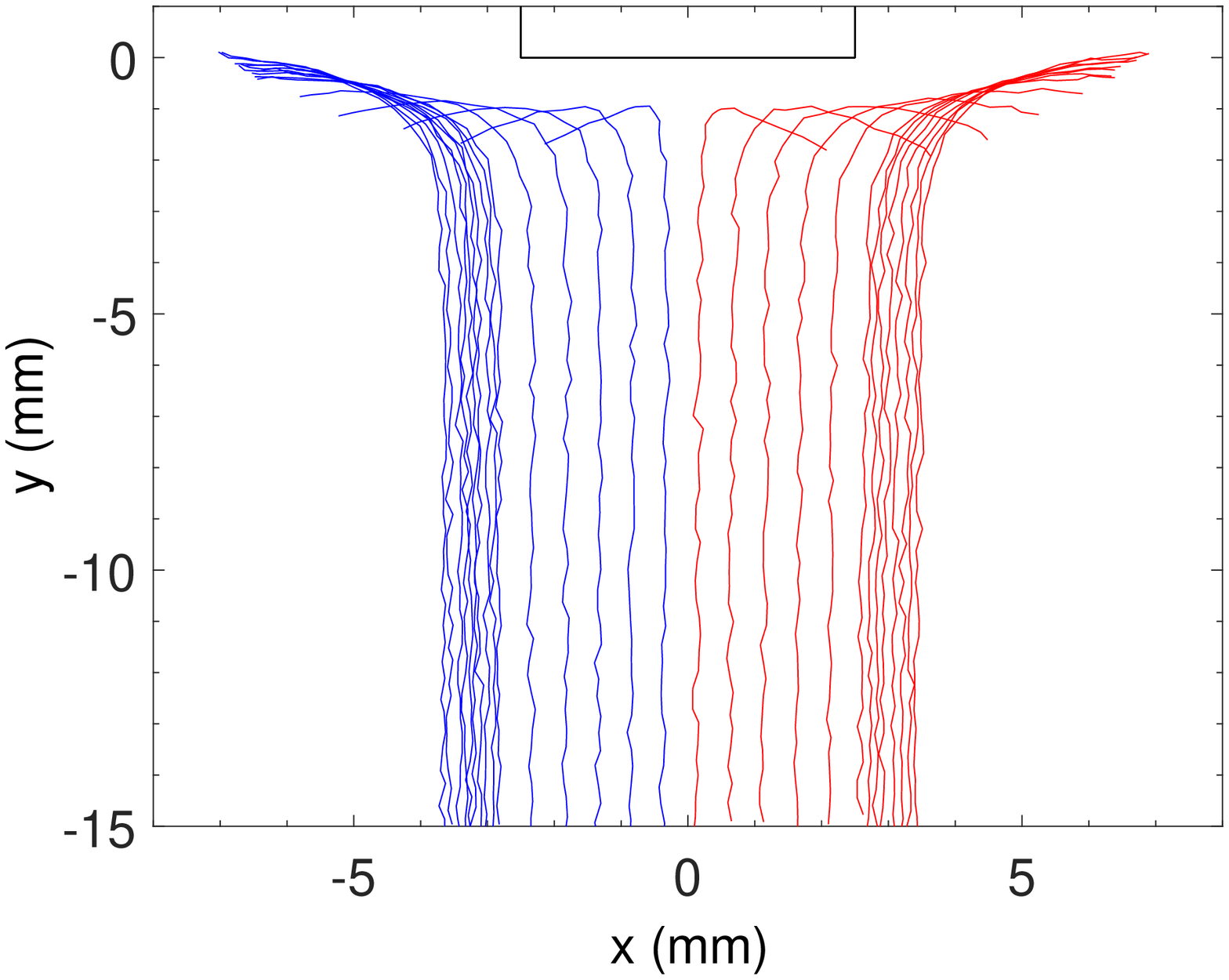,width=8.6cm}}
\caption{(Color online) Tracks reflected from the central part of the barrier. Note the symmetry around the center and that the approach is perpendicular to the barrier while the reflected tracks are strongly attracted to the nearest slit.}
\label{MiddleBlock}
\end{figure}

This is displayed in Fig.~\ref{MiddleBlock}. As seen, the in-going part of all tracks is nearly perpendicular on the barrier and first starts to deviate a few mm before impact. The outgoing part, however,  deviates immediately towards the nearest slit with an abrupt change in direction at the center of the barrier. Thus the movement of the droplet seems to be much more dependent on the form of structures behind than on structures in front of it.

This would explain why the tracks passing through a slit first seem to turn at a distance of more than a wavelength behind the barrier as seen in e.g. Fig.~\ref{ComplexPic} since the waves transmitted backwards from the droplet have to catch up again after reflection from the barrier. Any waves passing through the slit not used by the particle should have established a wave pattern on the far side long before the particle and its associated wave field appears in that region, the Faraday wave velocity being an order of magnitude higher than the particle velocity, and therefore would be expected to influence the tracks at once as the droplet emerges from the slit.  A more detailed explanation would require a closer study along the lines of Ref.~\cite{Andersen3}. 

The natural conclusion drawn from these different experiments is that the deflections observed in the previous figures are due to the reflection of waves from the outlet side of the unused slit interfering with the droplet-wave entity. All in all, it seems certain that the observed asymmetry is caused by interference effects purely originating in the space behind the barrier and is not connected to waves passing through the slit not used by the droplet (see also App. D for an overview of double-slit experiments).

\begin{figure*}
\includegraphics[width=17cm]{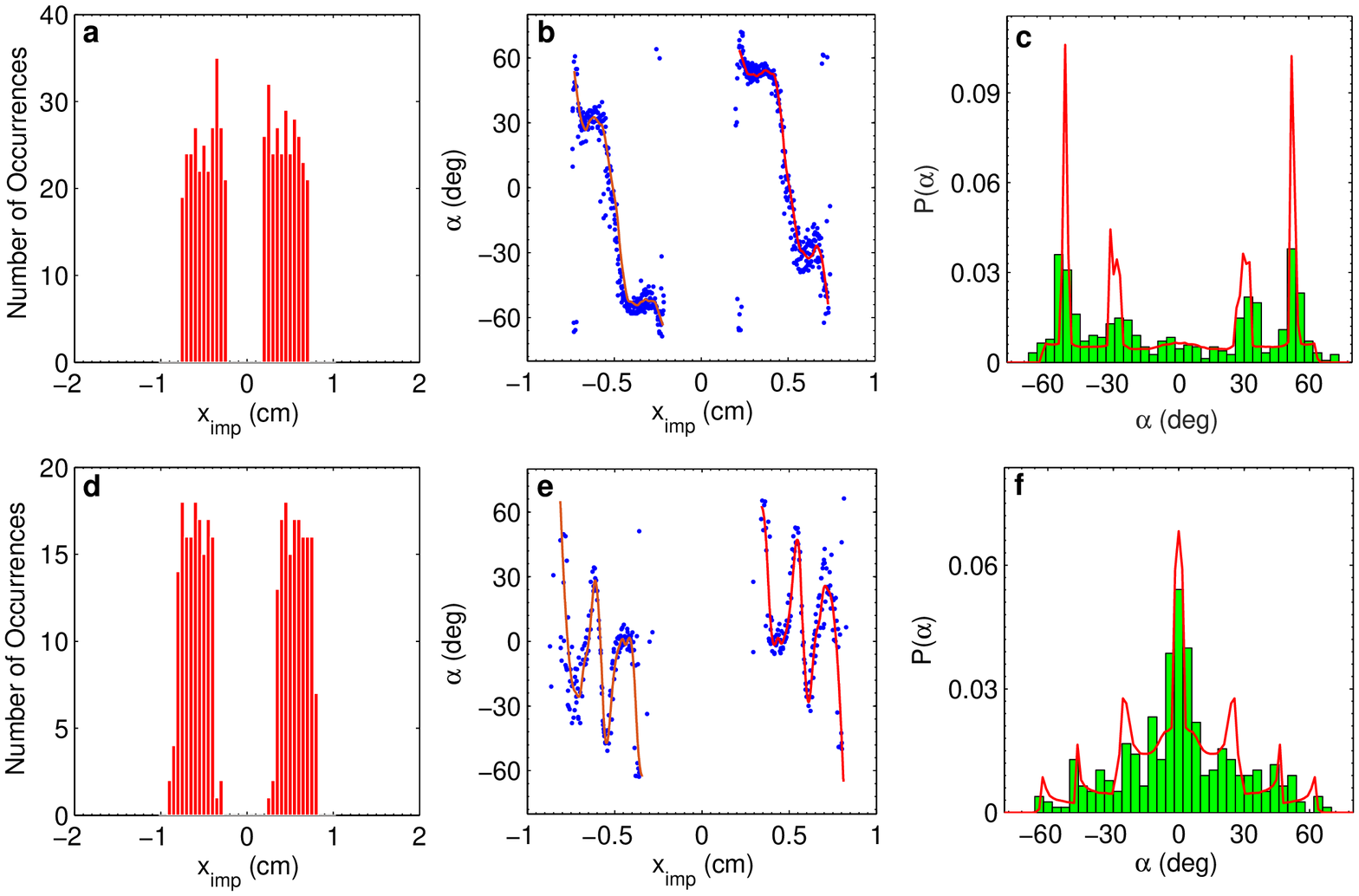}
\caption{(Color online) Statistics of double-slit experiment. Distribution of impact parameters x$_{imp}$ over double-slit (a) 5 mm slit, 557 tracks; (d) 7.5 mm slit, 331 tracks. (b) and (e) Corresponding $\alpha$ versus x$_{imp}$ ($\cdots$) with fit ($-$). (c) and (f) P($\alpha$) ($-$) from fit, normalized histogram showing experimental distribution of $\alpha$.  5 mm slit: v$_{imp}$ = 11.1 mm/s, $\gamma/\gamma_F = 0.999$,  h$_1$= 0.71 mm. 7.5 mm slit: v$_{imp}$ = 9.3 mm/s, $\gamma/\gamma_F = 0.997$,  h$_1$= 0.63 mm.}
\label{S18og19a}
\end{figure*}

Even then it is interesting to see what kind of density histograms one can obtain in these cases. In Fig.~\ref{S18og19a} we display an example (upper row) for the 5mm double-slit with a total of 557 tracks together with an example (lower row) for the 7.5 mm slit with a total of 331 tracks. Also shown are the calculated probability density distributions together with the actual measured normalized count distributions.

In the first column of Fig.~\ref{S18og19a} ((a) and (d)) is shown the distributions of input tracks leading to passage of the two slits, in both cases displaying a fairly homogeneous impact distribution. 
As seen from (b) and (e) we do find evidence for interference effects. The distribution (c) for the 5 mm slit shows no resemblance to the quantum mechanical interference formula. However, the distribution (f) found for the 7.5 mm case does have some resemblance to the formula although consideration of (e) shows this to be just a mere coincidence. From these measurements it is also clear that even with 557 tracks we still do not have good enough statistics to really pin down the underlying density distribution function. 

\noindent
\subsection{On the possible existence of chaos}
\begin{figure*}
\includegraphics[width=17cm]{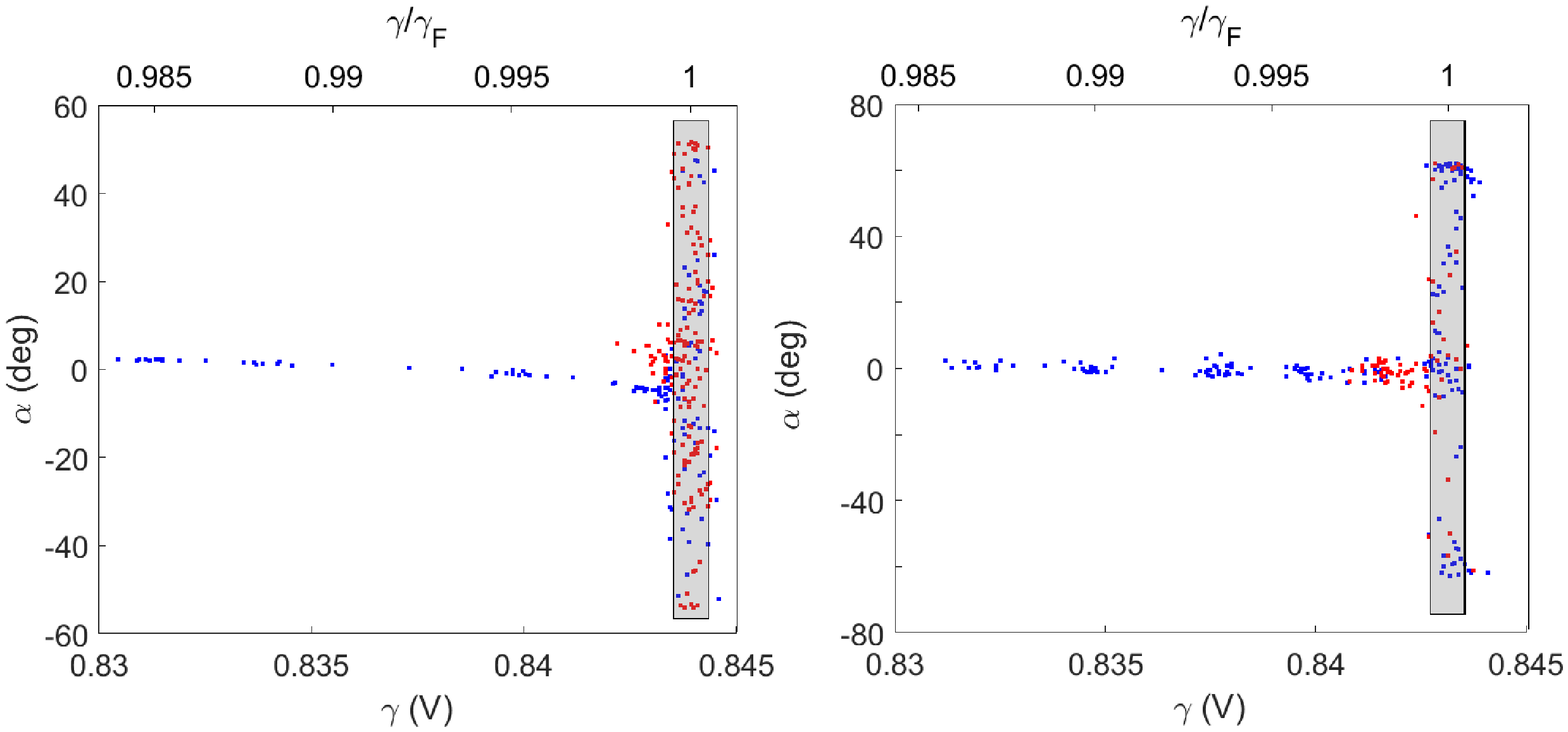}
\caption{(Color online) Checking for chaos. $\alpha$ versus measured drive acceleration for fixed x$_{imp}$. Note that the x-axis coverage is less than half of that in Pucci {\it et al.} \cite{BushNew}. Left panel: Double-slit. Overlay of two measuring series with a shift of 0.06 mm in x$_{imp}$. 285 tracks, v$_{imp}$ = 10.0 mm/s for the whole range. Right panel: Single-slit.  Overlay of two measuring series with fixed x$_{imp}$. 266 tracks, v$_{imp}$ = 8.2 mm/s for the whole range. For all w = 7.3 mm and h$_1$ = 0.62 mm. The gray rectangles signify the uncertainty in the determination of $\gamma_F$.}
\label{chaos}
\end{figure*}

In this section we first describe results obtained using either a 7.3 mm single- or a 7.3 mm double-slit. Following that we show results obtained with a single-slit of width 14.7 mm, i.e. the same size as used by Pucci  {\it et al.} \cite{BushNew}. In Fig.~\ref{chaos} we display a collection of tracks with fixed impact parameter obtained by slowly increasing the acceleration.  The left panel shows a result for the double-slit and the right panel a result for the single-slit. 

As shown we do not observe the gradual fanning out of $\alpha$ with increasing acceleration, that is present in the corresponding figure by  Pucci {\it et al.} \cite{BushNew} (their Fig. 4), but rather an abrupt change in distribution at the critical acceleration $\gamma_F$.

However, by inspection of the first frames in each recording we find, that the reflected images of the LED's show that nearly all of the deviating tracks inside the gray zone around $\gamma_F$  are obtained while actually surpassing the critical acceleration.

\begin{figure*}
\includegraphics[width=17cm]{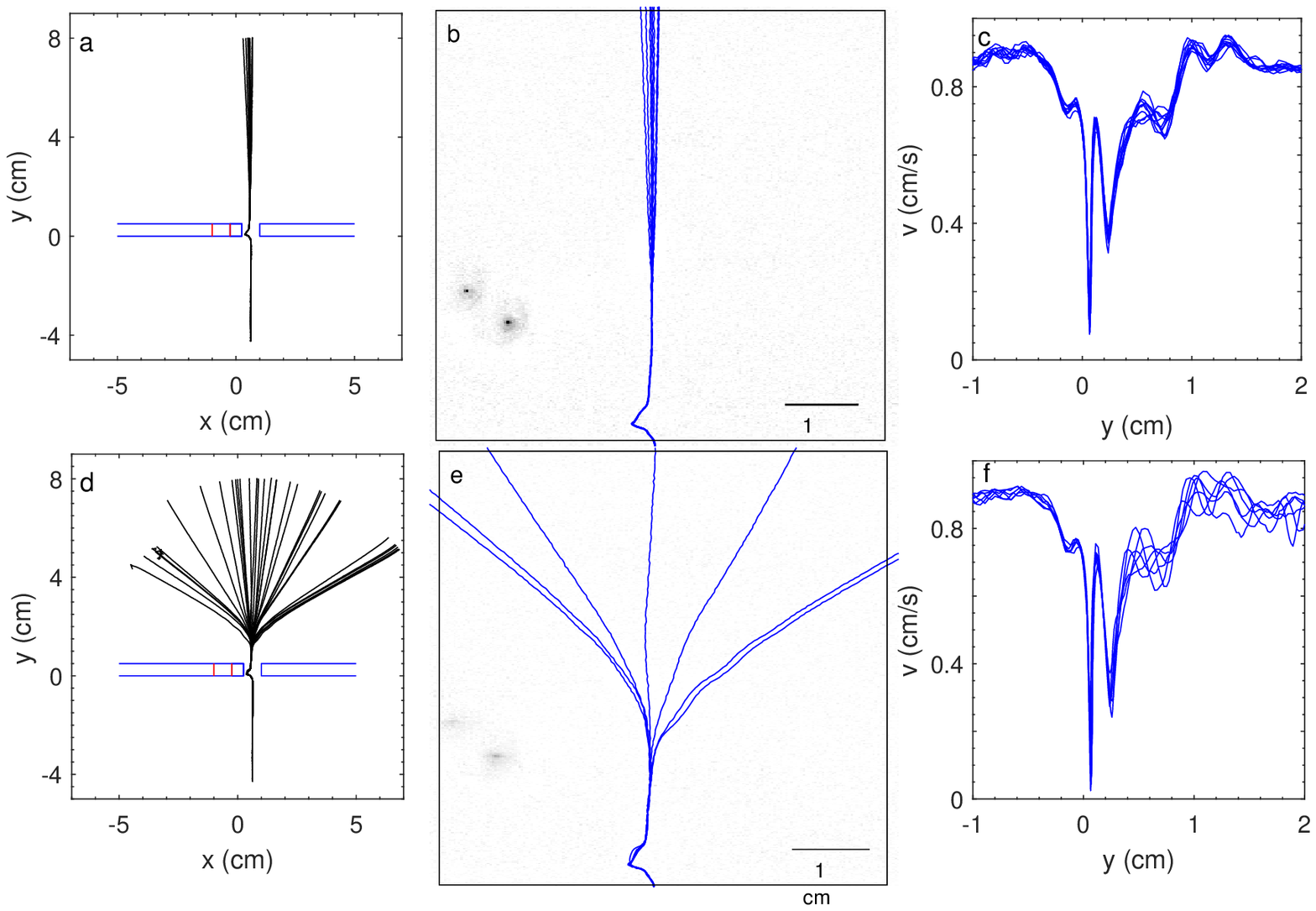}
\caption{(Color online) Upper row: (a) 10 Tracks from the sub-critical region of the single-slit experiment in  Fig.~\ref{chaos} ($\gamma/\gamma_F = 0.999$).  (b) Enlargement  of the region just above the slit showing also that the reflections of the diodes taken from the starting frames are sharp, i.e. that no waves are present. (c) Velocity plot in central region. Lower row: (d) Tracks from the gray regime at $\gamma_F$. (e) Enlargement of 7 tracks chosen to cover the span in $\alpha$ with most showing wiggling. The reflections of the diodes are smeared showing that waves are already present in the region above the slit when the droplets start. (f) Velocity plot of the tracks in (d) showing more noise due to the pre-existing waves than present in (c).}
\label{combined}
\end{figure*}

These features are demonstrated in Fig.~\ref{combined} with tracks used in  Fig.~\ref{chaos}(b). In (a) we display 10 tracks taken at $\gamma/\gamma_F$ = 0.999. In (b) we display a blown-up picture of the tracks overlaid on the combined inverted starting frames. The reflections of the diodes show up as sharp points here. In contrast we show in (d) tracks from the regime where the angles are spreading out. 7 tracks, representative of the spread in $\alpha$, are presented in (e) in a similarly blown-up picture. The measured drive amplitude is within the uncertainty equal to $\gamma_F$. Most of the tracks display strong visible wiggles or abrupt changes in direction and are therefore presumably obtained at values above  $\gamma_F$. Also, the diode reflections are now blurred. In (c) and (f) the respective velocity plots are presented. The difference between these plots is small and mostly shows that more noise, presumably due to the pre-existence of waves, is present for $\gamma \approx \gamma_F$.  

\begin{figure*}
\includegraphics[width=17cm]{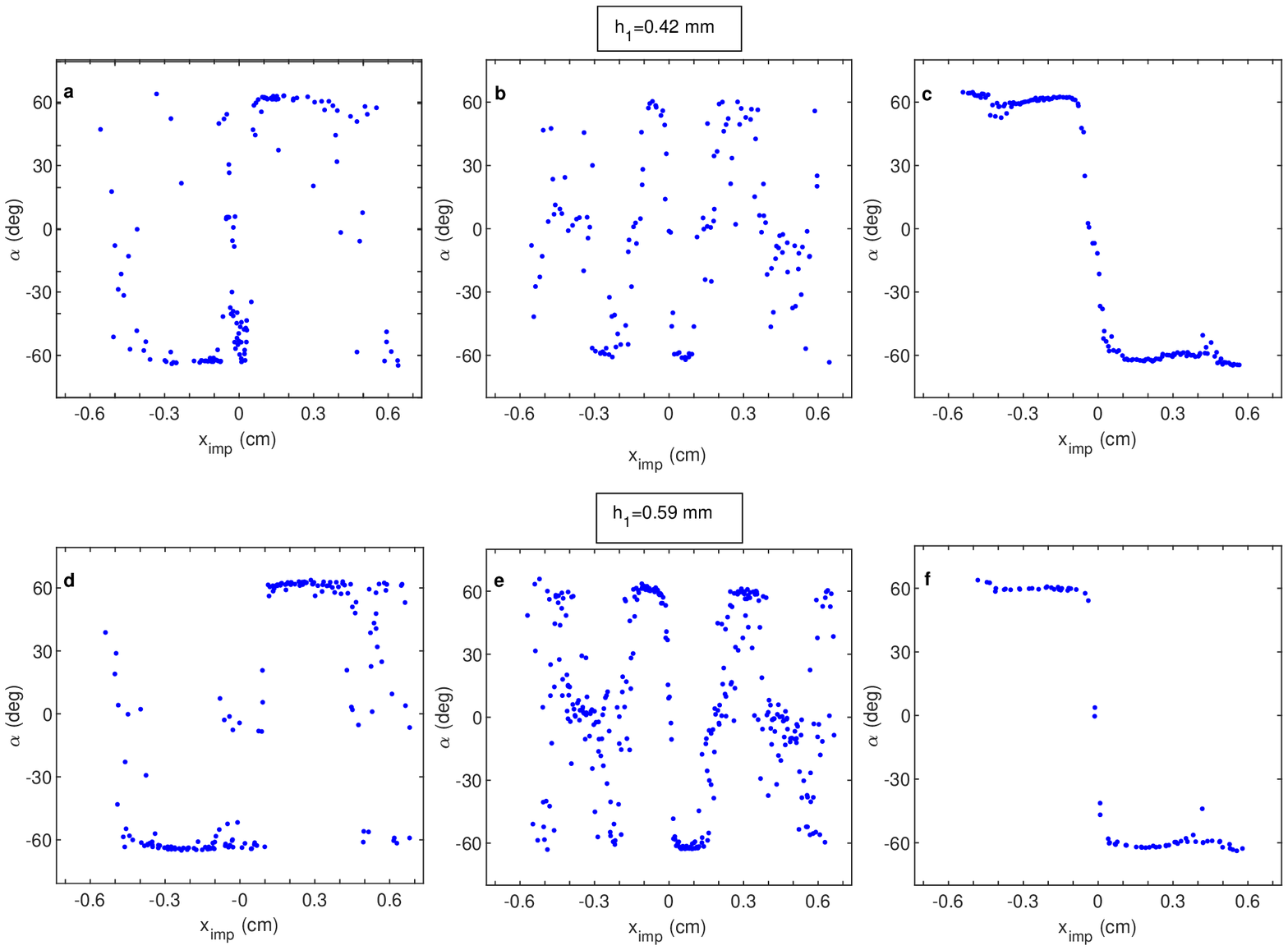}
\caption{(Color online)  $\alpha$ as function of x$_{imp}$ for a 14.7 mm single-slit. Top row $h_1$ = 0.42 mm and bottom row $h_1$ =0.59 mm.  Top row: (a)  187 tracks, v = 6.92 $\pm$ 0.04 mm/s, D = 0.61 $\pm$ 0.01 mm. (b)  136 tracks, v = 7.41 $\pm$ 0.02 mm/s, D = 0.63 $\pm$ 0.01 mm. (c) 140 tracks, v = 8.02 $\pm$ 0.04 mm/s, D = 0.66 $\pm$ 0.01 mm. Bottom row: (d) 163 tracks, v = 6.62 $\pm$ 0.06 mm/s, D = 0.60 $\pm$ 0.01 mm. (e) 281 tracks, 7.34 $\pm$ 0.03 mm/s, D = .61 $\pm$ 0.03 mm. (f) 91 tracks, v = 7.87 $\pm$ 0.03 mm/s, D = 0.65 $\pm$ 0.01 mm. For all cases $\gamma/\gamma_F$ =0.998. Note that the structures seen at $h_1$ = 0.42 mm are repeated for $h_1$ =0.59 mm.}
\label{147chaosExp}
\end{figure*}

We now turn to the measurements made with the 14.7 mm slit. In  Fig.~\ref{147chaosExp} we display $\alpha$ as function of x$_{imp}$ for values of height over the barrier $h_1$ = 0.42 mm and 0.59 mm, with three different values of drop-sizes per value of $h_1$. For all the cases $\gamma/\gamma_F$ = 0.998. 
We see that the structures present in the measurements taken at  $h_1 = 0.42$ mm are repeated at $h_1 = 0.59$ mm although for slightly lower velocities. Note that  Fig.~\ref{147chaosExp}a corresponds to the example shown in Pucci  {\it et al.} \cite{BushNew} (their Fig. 5). Also, that we see a complete flip of signs in the $\alpha$-distribution from  Fig.~\ref{147chaosExp}a to Fig.~\ref{147chaosExp}c and from Fig.~\ref{147chaosExp}d to Fig.~\ref{147chaosExp}f within a small range of velocities.

\begin{figure}
\includegraphics[width=8.6cm]{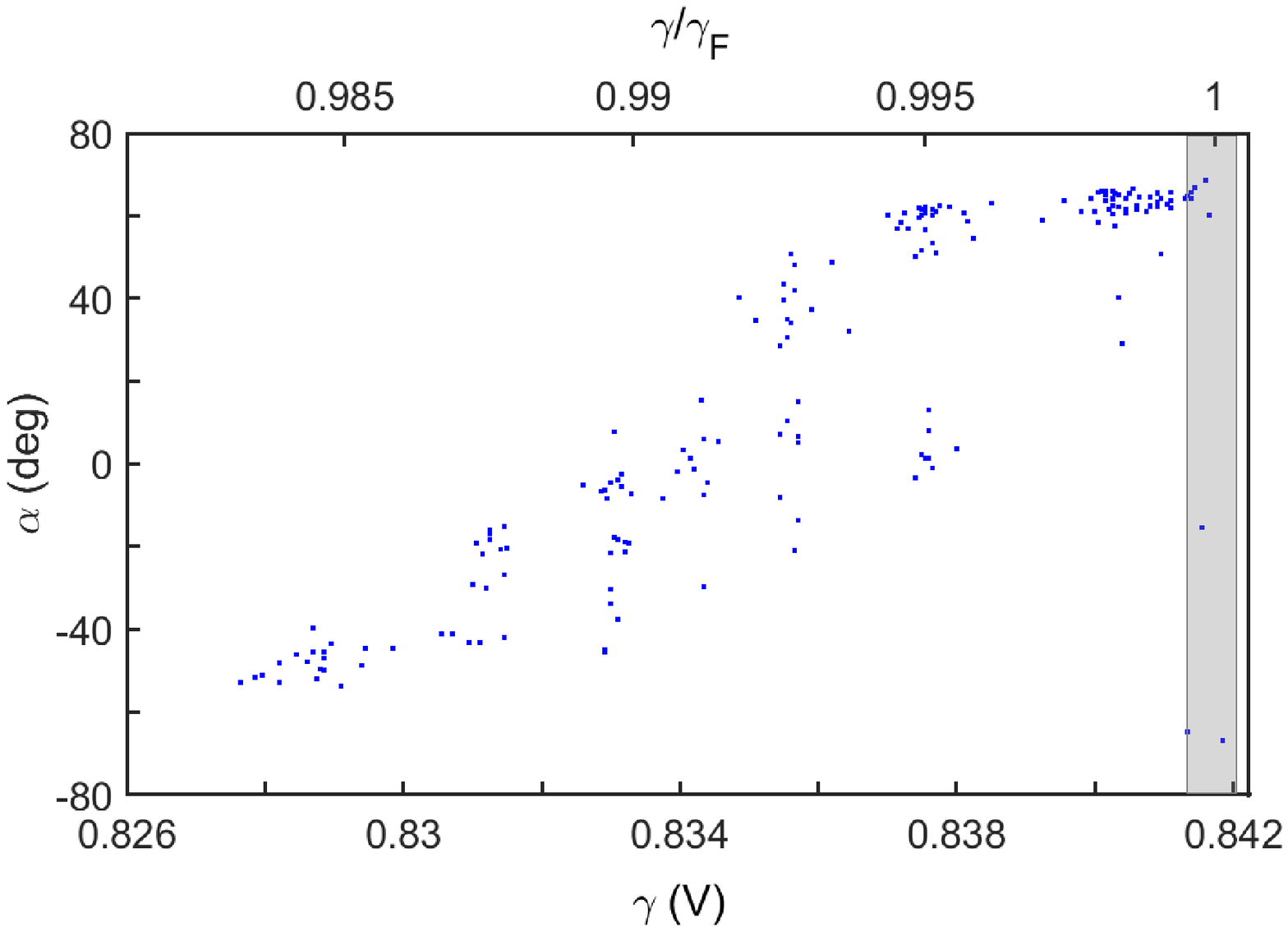}
\caption{(Color online)  $\alpha$ as function of $\gamma$ for 14.7 mm single-slit with $x_{imp}$ fixed at 2.6 mm, 186 tracks, D = .61 $\pm$ 0.01 mm. v ranging from 6.65 $\pm$ 0.05 mm/s at $\gamma/\gamma_F$ = 0.985 to 7.10 $\pm$ 0.05 at $\gamma/\gamma_F$ = 0.999. $h_1 = 0.42$ mm. Thus the parameters are close to those used by Pucci {\it et al.}. The gray rectangle signifies the uncertainty in $\gamma_F$.}
\label{FE86Exp}
\end{figure}    
In  Fig.~\ref{FE86Exp} we have furthermore repeated the measurement performed by  Pucci {\it et al.} for fixed $x_{imp}$ and displayed in their Fig. 4e. Here $x_{imp}$ is chosen to be  $\sim 2.6$ mm (corresponding to $x_{imp}/L \sim 0.18$). While there is an intermediate regime where the angular response spreads out, obviously the range in $\alpha$ values narrows sharply when $\gamma$ gets close to $\gamma_F$. Although the data shown in  Fig.~\ref{14.7mmAmplVar} are obtained at a different value of h$_1$, they provide a clue to what goes on. At $\gamma/\gamma_F \sim 0.985$, P($\alpha$) is flat around $x_{imp} \sim 2.6$ mm giving a plateau of $\alpha = -60^{\circ}$. As $\gamma$ is increased, a peak starts to grow to towards positive values of $\alpha$. Between $\gamma/\gamma_F \sim 0.995$ and 0.998, this peak grows rapidly and then saturates and broadens out to give a plateau now at $\alpha =  +60^{\circ}$. If  $x_{imp}$ is situated so the flank of the peak is sweeping by and $\alpha$ ends up on the saturated plateau, we get the picture shown in Fig.~\ref{FE86Exp}. However, a small change in $x_{imp}$ and/or D will keep $\alpha$ on the flank of the peak. This could be a possible explanation for the result found by Pucci {\it et al.}. As already mentioned there is a rapid change in the $\alpha$ response in this range of drop size. Taken together with the clear and reproducible structures observed in Fig.~\ref{147chaosExp}, we must conclude that these measurements do not support the notion of a chaotic regime.

To conclude this section we note that none of the measurements in our rather extensive coverage of the regime of chaotically bouncing drops at high $\gamma$ values support the notion of chaos. Also that none of the measurements shown in Figs.~\ref{5mmDepthVar}, \ref{5mmVelVar}, \ref{7.5mmVelVar}a,d,and g, \ref{7.5mmAmplVar}, and \ref{Ang_dist} would have been possible if chaos was present.

\noindent
\subsection{Strange tracks}
\begin{figure*}[th]
\includegraphics[width=17.2cm]{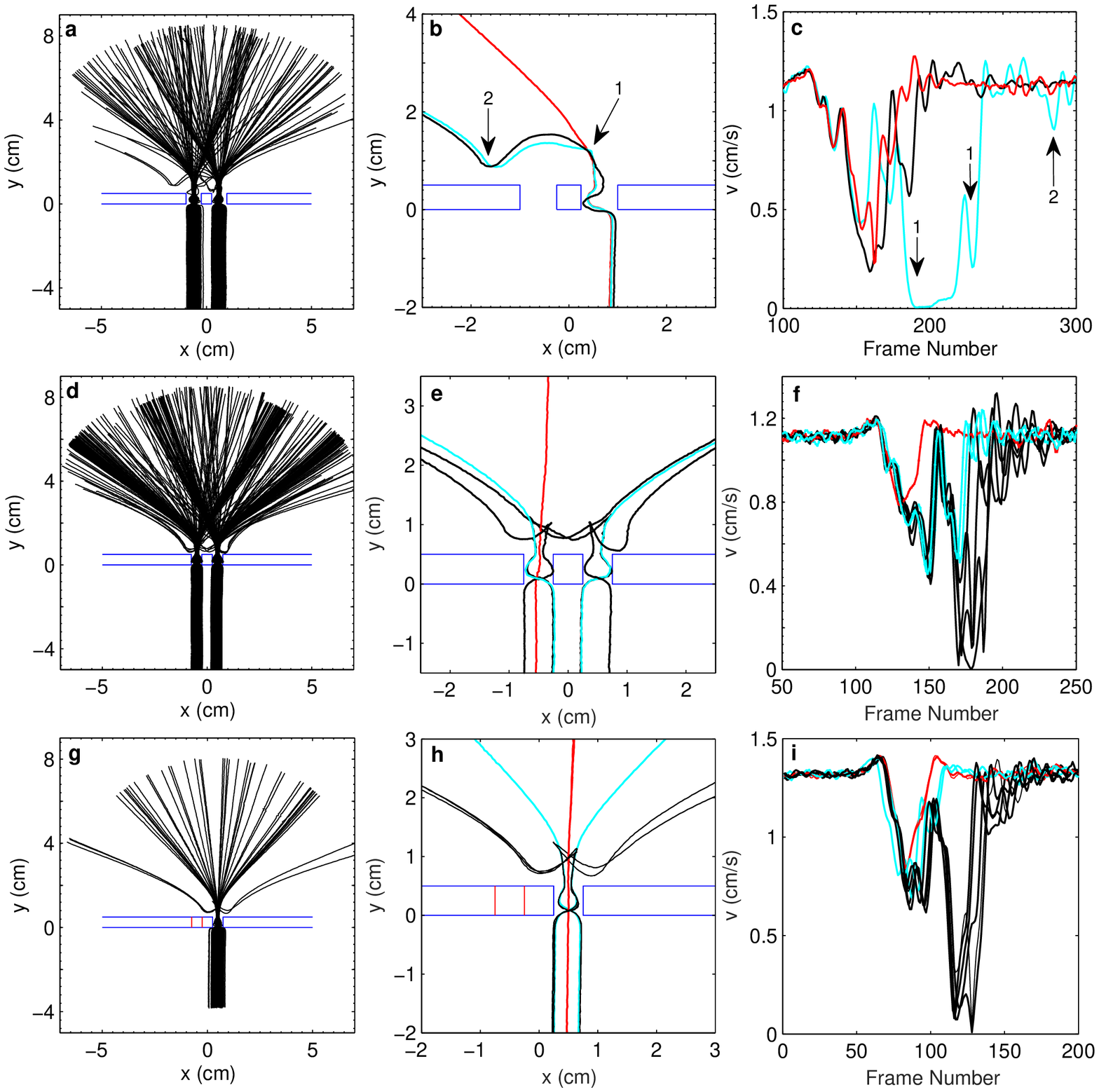}
\caption{(Color online) Examples of strange tracks. Top row: 7.5 mm double-slit, from  Fig.~\ref{Ang_dist}a. Center row: Double-slit, from  Fig.~\ref{Block}a.  Bottom row: Single-slit, h$_1$ = 0.81 mm. Left column: Tracks. Center column: back-bending tracks isolated, for comparison we also show some tracks behaving as normal. Right column: droplet velocity v of tracks in (b) as function of frame number. Arrows in b) and c) are corresponding.}
\label{weird}
\end{figure*}

Just before the end of the range of possible values of the impact parameter  x$_{imp}$ we often encounter strange looking tracks. Examples from both the 7.5 mm and the 5 mm slits are shown in Fig.~\ref{weird}. 
In the top row we present tracks from a scan over possible impact parameters for  a double-slit experiment with slit size 7.5 mm. Several back-bending tracks are discernible in Fig.~\ref{weird}a. In b) we have isolated three tracks, two of which are clearly of the same type as those observed by Pucci {\it et al.} (their fig.~7a), where the tracks are attracted to the unused slit and then reflected from the barrier. One track follows a rather smooth curve whereas the other exhibits a sharp bend marked by the arrow 1. For comparison we also show a track behaving as normal. To further investigate what goes on we show in c) the corresponding velocities v(t) of the droplet along the tracks. The normal track shows the usual behavior with a dip in v(t) inside the slit. The smooth back-bending track has a second dip at the arrow 1 but recovers fast while the track with the sharp bend spends a comparatively long time in the immediate vicinity of the bend and has a new dip in v(t) at the arrow 2 right after being reflected from the barrier. All in all the behavior of the velocities points to a complicated interaction of the droplet with a wave pattern being influenced by both slits.   

However, a careful look at Fig.~\ref{weird}a reveals the existence of some other complicated structures. These structures are even more prominent in Fig.~\ref{weird}d that shows the tracks obtained for a 5 mm double-slit (same as Fig.~\ref{Block}a). In Fig.~\ref{weird}e we have isolated four of the back-bending tracks that exist in a near symmetric fashion but are also showing two nearby tracks and a central track for comparison. These back-bending tracks seem to be related to the track with the sharp bend displayed in Fig.~\ref{weird}b but instead of turning towards the unused slit they back-track and are then reflected by the barrier.  In Fig.~\ref{weird}f  we show the corresponding time-series of the velocity of the droplet. The same pattern as is seen for the 7.5 mm  double-slit is true here, with also the neighboring tracks coming close to a complete stand-still. The synchronization of the oscillations in v(t) again points to interaction with a wave pattern complicated by the intricate boundary conditions.  

The tracks in  Fig.~\ref{weird}g are part of a measurement where x$_{imp}$ has been swept back and forth across a single 5 mm slit. Of the five back-bending tracks isolated in h) the first left going is recorded at the beginning of the time series and the last two left going at the end. As seen the tracks are reproducible, leaving no doubt of their causal nature. We also display two neighboring  tracks that proceed as usual for comparison, showing that a tiny difference in impact parameter can result in widely different outcomes. In Fig.~\ref{weird}i we show the corresponding time-series of the velocities v(t) that also display a high degree of reproducibility. The back-bending tracks are seen to never completely recover after the velocity loss inside the barrier and instead come to a nearly complete stop before turning back. The nearby tracks also show some slowing down under the recovery but then speed up again. For comparison, we also show the behavior of two tracks passing through close to the center of the slit. 

To conclude this subsection, we have only seen the smooth back-bending track also observed by Pucci {\it et al.} for the wider slit where it might be related to the existence of the second slit. However, the back-tracking is observed for both single- and double-slits of both 5 and 7.5 mm sizes and thus seems caused by changes in the wave pattern due to the slit used. 

\noindent
\section{Conclusion:}
\begin{figure}[ht]
\includegraphics[width=8.6cm]{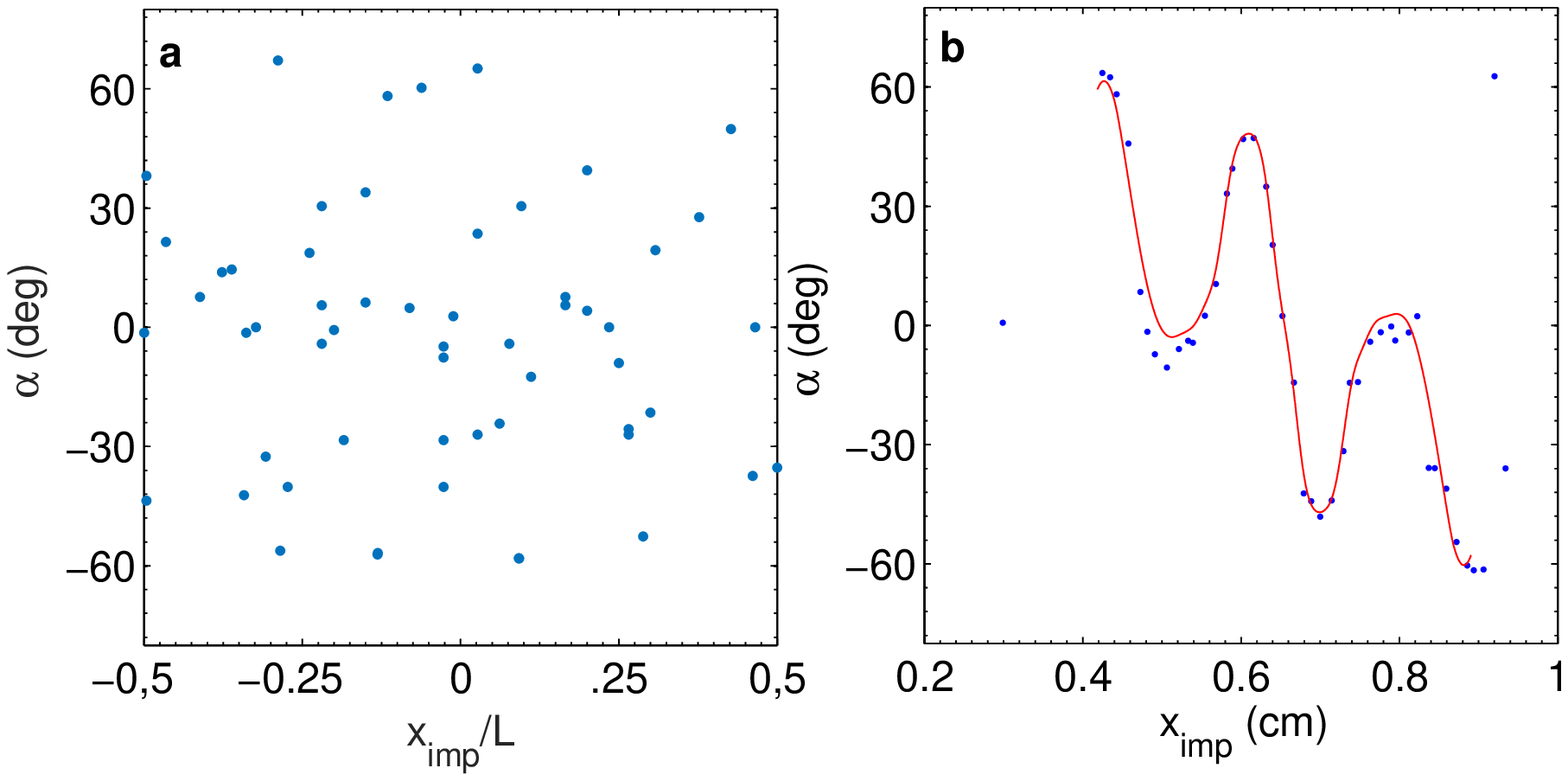}
\caption{(Color online) Deflection angle $\alpha$ versus x$_{imp}$. (a) Data taken from Couder et Fort  \cite{CouderFort} with x$_{imp}$ normalized with the slit width L. (b) Sample of our measurements, taken from Fig.~\ref{7.5mmAmplVar}b.}
\label{random}
\end{figure}

Inspired by the provocative paper by Couder and Fort \cite{CouderFort} we have undertaken a rather comprehensive investigation of walking droplets traversing both single- and double-slit configurations. A hermetically closed container and strict control of temperature, droplet size, and acceleration are found to be absolutely essential for a reproducible outcome. For a single-slit there are obvious interference effects due to the local interaction of the self-generated waves with the slit structure, giving rise to a fanning out of the outgoing tracks even for parallel input tracks. It is, however, also clear from our measurements that for accelerations below critical, the situation is strictly causal with a sharp correlation and no sign of the randomness claimed by Couder and Fort (see Fig.~\ref{random}).

While experimental parameters can be chosen so the resulting angular distribution P($\alpha$) does show some similarity to that of the quantum single-slit result, scanning the parameter space  reveals the existence of a zoo of radically different P($\alpha$)'s. In the limited regime investigated in detail by Pucci {\it et al.}  \cite{BushNew} they demonstrated that the interaction of walking droplets with slits was dominated by a wall effect resulting in two peaks at a limiting angle $|\alpha_{lim}|$ and possibly a central peak in the P($\alpha$) stemming from impact parameters in the mid-section of the slit. We do find that the limiting angular response $|\alpha_{lim}|$ never seems to exceed 70 $^{\circ}$. However, our more complete coverage of the chaotically bouncing droplet regime shows that the picture presented in Pucci {\it et al.} is  much over-simplified with a peak at $|\alpha_{lim}|$ not necessarily being the dominating feature in P($\alpha$). Also, a central peak might arise from values of $x_{imp}$ far from the center of the slit. For the larger slits e.g. Figs. ~\ref{14.7mmVelVar}, ~\ref{14.7mmAmplVar}, and ~\ref{147chaosExp} we find that the angular response in some cases sweeps back and forth in a wave-like fashion resulting in many peaks in P($\alpha$). Thus the system is fascinating in itself with wave-guided particles even if it is not quantum physics.

The double-slit experiments do show an extra interference effect. This was also found by Pucci {\it et al.}  \cite{BushNew} who however did not establish the cause. Nor were Andersen {\it et al.} \cite{Andersen3}  able to draw any firm conclusions concerning the cause from experiments, where they measured the height and distribution of the wave field. 

By blocking respectively the input or output sides of the slit not entered by the droplet, we have been able in our experiments to establish that the observed interference has nothing to do with waves traveling through the unused slit but is solely caused by back-scattering of the wave field on the outlet corners of the unused slit after a successful passage. 

An interesting idea is whether a chaotic situation might arise close to the critical acceleration that could lead to a quantum like randomness. This idea seemed to have some merit as Pucci {\it et al.}  \cite{BushNew} (see also ref. \cite{Harris}) indeed report that tracks having a fixed value of x$_{imp}$ do seem to fan out after passage of a slit in a seemingly chaotic fashion as the acceleration gets close to critical. We have investigated this regime carefully varying the droplet free speed from 6.4 to 10 mm/s, thereby covering most of the chaotically bouncing droplet region. The fluid height h$_1$ over the barrier has been varied from 0.4 - 1 mm,  and four different slit configurations investigated. However, nowhere do we find support for the existence of a chaotic situation before surpassing the critical acceleration. Instead we find reproducible structures as seen in Figs.~\ref{14.7mmVelVar} and \ref{147chaosExp}, that vary rapidly with small changes in $\gamma$ and droplet velocity (droplet size). These variations easily explain the fanning out in Fig.~\ref{FE86Exp} and in Fig. 4 by Pucci {\it et al.}. The fanning out comes when a steep flank in the angular response is passing the chosen value of $x_{imp}$ while changing the acceleration. It happens close to $\gamma_F$, but the fan can actually gather again to a narrow bunch as seen in Fig.~\ref{FE86Exp}. Thus the apparent randomness is most likely due to an extreme sensitivity  close to $\gamma_F$ to variations in experimental parameters, and is not really quantum like.  

The authors wish to thank A. Boisen and J.C. Hansen for help with electronics, Erik Hansen from DTU Physics for technical help, and A. Andersen, T. Bohr, and B. Lautrup for illuminating discussions. 

\noindent
\section{Appendix A}
The container is 20.2 cm outer diameter all black made in anodized aluminum. An inner basin of 18 cm in diameter is surrounded by a shelf 5 mm high and 7 mm broad for damping out waves at the rim. It has a built-in accelerator, which in later experiments was shortened, and a central groove spanning the entire interior where a movable barrier with a double slit can be placed (see Fig.~\ref{tray} for scaled details). A transparent plastic cylinder is placed inside the vessel and supports a glass lid sealed with a rubber gasket. The system has been analyzed and checked for resonances and deformations. The lowest resonance of the container is around 2.5 kHz, and the 4 g load gives a sub-micron maximum deformation of the bottom under working conditions. The lowest resonance of the glass lid is around 950 Hz, with the load giving around 2.5 $\mu$m deformation under working conditions.

The nozzle of the drop cannon enters through a hole in the plastic cylinder with the opening around the  nozzle sealed with a thin rubber membrane. Thus the interior of the vessel forms an airtight enclosure although a 3mm plug allows for diffusive pressure equalization. The vessel is supported on an aluminum cone mounted on a stainless steel cylinder that again is mounted on a LDS V400 exciter driven by an LDS PA500L-CE amplifier. This in turn gets its input signal from an Agilent 33220A Waveform Generator. At the base of the cone is also mounted a Bruel \& Kjær (Deltatron Type 4534-B-001) accelerometer. A tight fitting Teflon coated aluminum block surrounds and guides the steel cylinder in order to prohibit horizontal oscillations \cite{Harris and Bush}. Two accelerometers (ADXL 103CE) mounted on the vessel show that the magnitude of such sideways oscillations have amplitudes below the noise level of the accelerometers (i.e. below a few tens of a percent of the drive, see footnote \cite{ADXL}). The exciter is secured to a thick and heavy metal plate which is also the support of the guiding aluminum block. This plate is loaded with lead blocks and rests on special rubber vibration dampers.

The spectral purity of the vibration has also been checked under operational conditions. The peak of the first harmonic is always within 0.01\% of 80 Hz, and the second harmonic is found to be at least 34 db below the drive with all other higher harmonics substantially below. Thus the total harmonic distortion (THD) is at most 0.02. The flat noise floor is alway $\sim$ 85 db below and does not vary with time. The long time stability is given by the precision by which we can keep $\gamma$ constant ($\pm 0.05$ \%). Thus, even with a simpler system of guiding, our setup is comparable to that of Ref.~\cite{Harris and Bush}.   

The aluminum cone is surrounded by a table on which is mounted the support for the drop cannon and a step motor for external adjustment of the barrier position. A cylinder covered with thermal isolation material and with a glass window in its top plate is placed on the table around the vessel for better temperature stability. Slits in this cylinder covered with light diffusing material allow for sideways illumination of the vessel using an LED array. The temperature of the vessel is measured and kept at the desired operating value by a feedback system using resistors mounted at the bottom of the vessel. Likewise is the temperature of the steel tube kept at a slightly lower temperature using a water cooled Peltier element in order to isolate from the heat generated by the exciter, hereby providing temperature stabilization for the accelerometer, while the room  temperature is controlled using an electric heater. The various temperatures are measured using calibrated thermistors and the control implemented by an autonomously running Arduino UNO R3 board. Finally the temperature of the glass lid is monitored and held from deviating more than 0.5 K from the oil temperature.  

\begin{figure*}[ht]
\centering
\includegraphics[width=17.2cm]{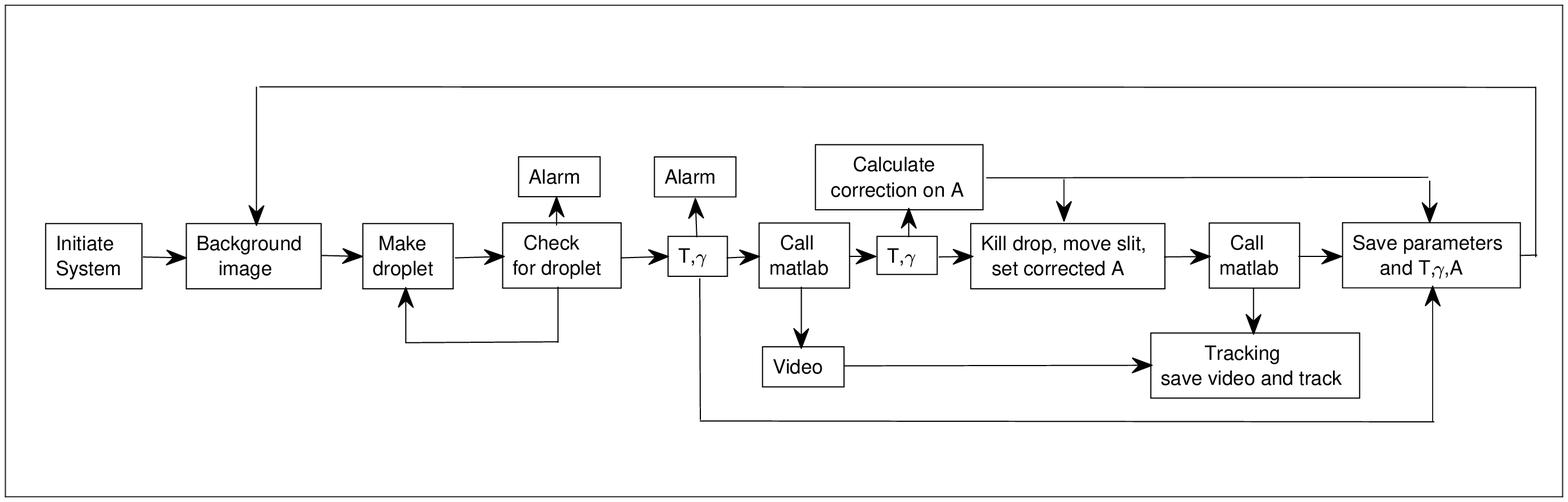}
\caption{Flow-diagram  showing the main features of the LabVIEW program. A is the amplitude supplied by the signal generator, while T represents all four temperatures measured.}
\label{flowdiagram}
\end{figure*}

The entire experiment is automated, controlled by a LabVIEW program running in a while loop (see Fig.~\ref{flowdiagram}). The drive amplitude can be locked to produce a certain percentage of the critical acceleration. A droplet is made by the cannon and a still picture used to check for success. If no droplet is found the drive amplitude is lowered for a short period to ensure that any drop that may just be out of the field of view is killed. A new droplet is created after a wait period long enough to reach a stable situation again and the process repeated. If a droplet is found a short control measurement of the amplitude is performed before a MATLAB program is called that controls the camera. A predetermined number of frames is recorded. A new measurement of relevant temperatures and the acceleration is done and the drive amplitude then shortly lowered to kill the droplet. In the down period the barrier is moved if necessary. Then a second MATLAB program is called that tracks the droplet and stores the track in a text file and the frames in an AVI file, where-after the whole process is repeated. At the end of a run information about all relevant parameters such as frequency, amplitude of drive, temperatures and accelerations, landing position for the drop etc. is stored in a separate text file. 

The drop cannon is inspired by the droplet-on demand design by Harris {\it et al.} \cite{Harris2} but here shooting droplets out horizontally. The LabVIEW generated trigger signal is passed to a flip-flop circuit that generates a trigger signal phase-locked to the drive. This trigger signal is used to gate a BNC model 565 pulse delay generator that in turn generates the final trigger signal for the cannon. Correct choice of the phase-delay leads to a near perfect 100 \% success rate in generating a droplet, the size of which can within some limits be controlled by setting the amplitude and pulse length of the trigger pulse.

Three different double-slit barriers of height and depth 5 mm but varying slit width were used, while the central part of the barrier was kept fixed at 4.7 mm. A single-slit barrier with slit width 14.7 mm was also used. Plugs of different width can be inserted in one of the slits through the hole for the drop cannon in order to close a slit on either the entrance side or the outlet side without removal of the top-plate. The barriers can be moved from the outside by the LabVIEW program during a run by a step-motor that is programmed to engage for the move and then disengage. Thus a controlled sweep of the impact parameter can be made in steps as small as 0.05 mm.

\noindent
\section{Appendix B}
\begin{figure}[ht]
\includegraphics[width=8.6cm]{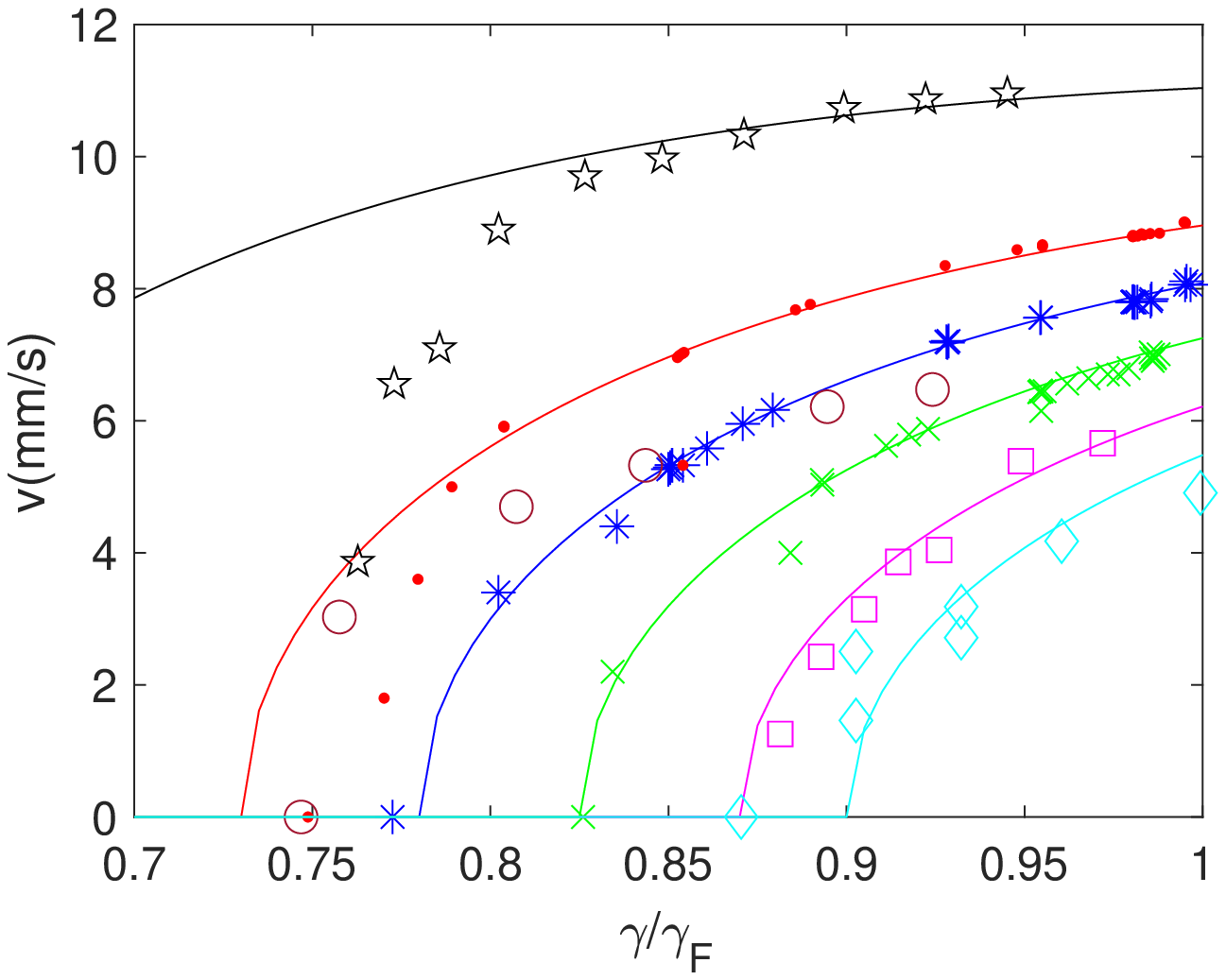}
\caption{(Color online) The free speed v as function of the drive amplitude for 6 different droplet sizes. All are fitted to a simple empirical expression Eq.~\ref{SizeVel}. Our measurements: D =0.63 mm (x), D = 0.66 mm ($\ast$), D = 0.69 mm ($\cdot$). From Mol\'{a}$\check{\rm c}$ek and Bush \cite{molacek}: D = 0.62 mm ($\square$), D = 0.76 mm ($\star$). From  Proti\`{e}re  {\it et al.} \cite{protiere} : D = 0.56 ($\diamond$), D = 0.70 ($\circ$).}
\label{V-Amp}
\end{figure}

Fig.~\ref{V-Amp} shows the dependence on drive amplitude of the free speed for some slower droplets in the regime of chaotically bouncing walkers, while Fig.~\ref{V-D} shows the connection between droplet size and free speed at $\gamma_F$. 
Unfortunately no simple analytic expression exists. However, using Eq.~\ref{SizeVel}, Durey \& Milewski \cite{Durey} Eq.~5.23 (see also Proti\`{e}re  {\it et al.} \cite{protiere} Eq.~5), with the prefactor $\Gamma_2$ empirically determined to be $\Gamma_2^{-½} =26 (1-x/3)/\sqrt{x}$, the measurements can be nicely fitted to a simple expression, with D = 0.70 mm by  Proti\`{e}re  {\it et al.} \cite{protiere} as the sole exception. 
\begin{equation}
v(x)= \Gamma_2^{-½} \sqrt{x-x_w}
\label{SizeVel}
\end{equation}
with $x = \gamma/\gamma_F$, $x_w = \gamma_w/\gamma_F$ with $\gamma_w$ denoting the drive at the onset of walking for droplets entirely in the regime of chaotically bouncing walkers. 

\begin{figure}[ht]
\includegraphics[width=9cm]{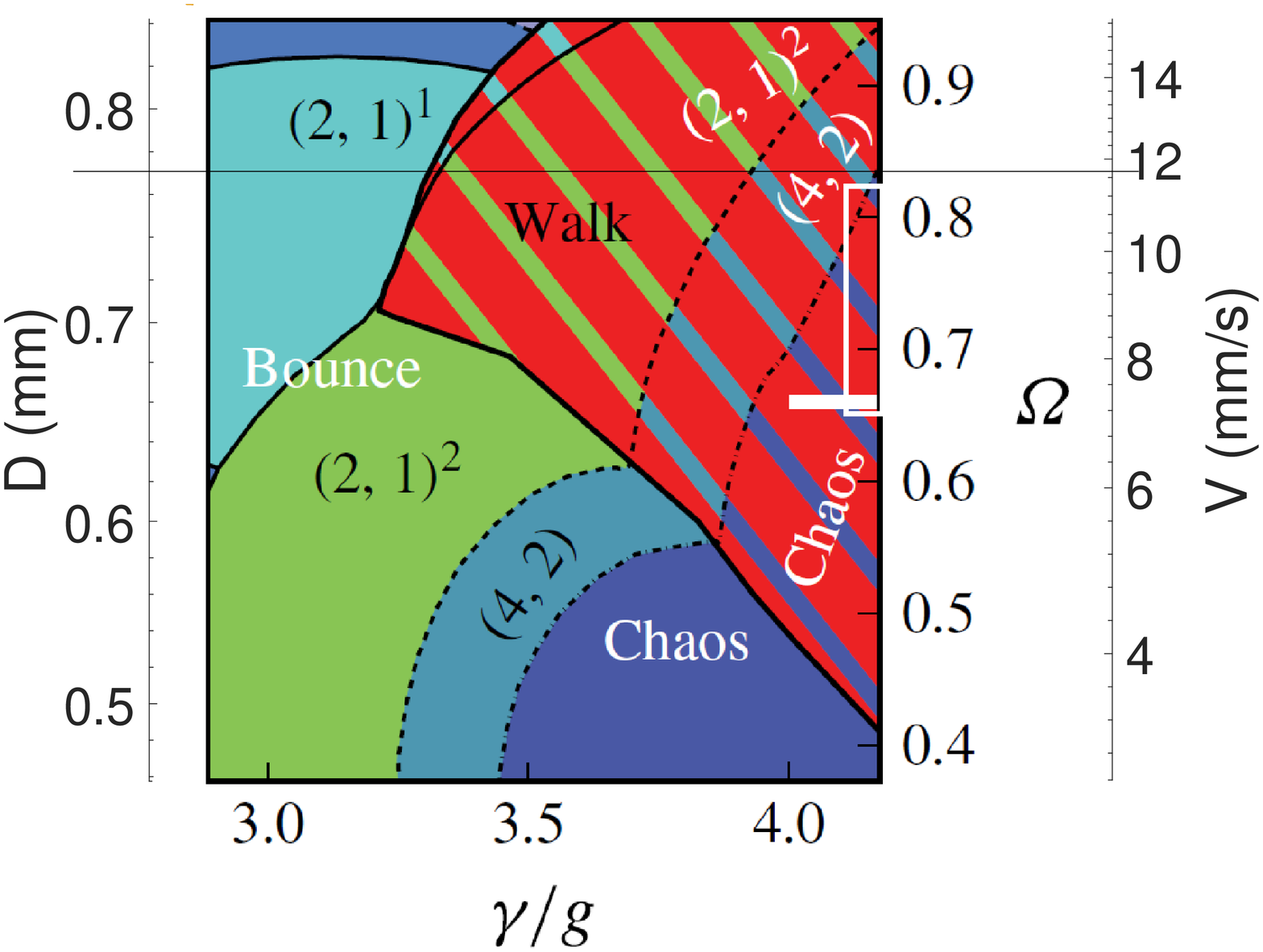}
\caption{(Color online) Phase-diagram taken from Pucci  {\it et al.} \cite{BushNew} showing the regime investigated in detail by these authors (white rectangle) and augmented with axes showing the corresponding droplet size and critical velocity. The white bordered rectangle shows the regime covered in this work, while the horizontal line shows the upper limit for the chaotic bouncing regime.}
\label{PhaseDiagram}
\end{figure}
Using the definition $\Omega = 2\pi f \sqrt{\rho D^3/8\sigma} $ we have added a droplet-size axis to the phase diagram \cite{WindWillassen,Pucci,Bush2015} as presented by Pucci  {\it et al.} \cite{BushNew}. We have then used the fitting curve h(x) from Fig.~\ref{V-D} to add an axis showing the corresponding critical free speed (Fig.~\ref{PhaseDiagram}). Furthermore we have added a white-bordered rectangle showing the parameter-region explored in detail in the present work. Nearly all our reported results are obtained from inside the regime of chaotically bouncing walkers.

\noindent
\section{Appendix C}
\begin{figure}[ht]
\centering
\includegraphics[width=8.6cm]{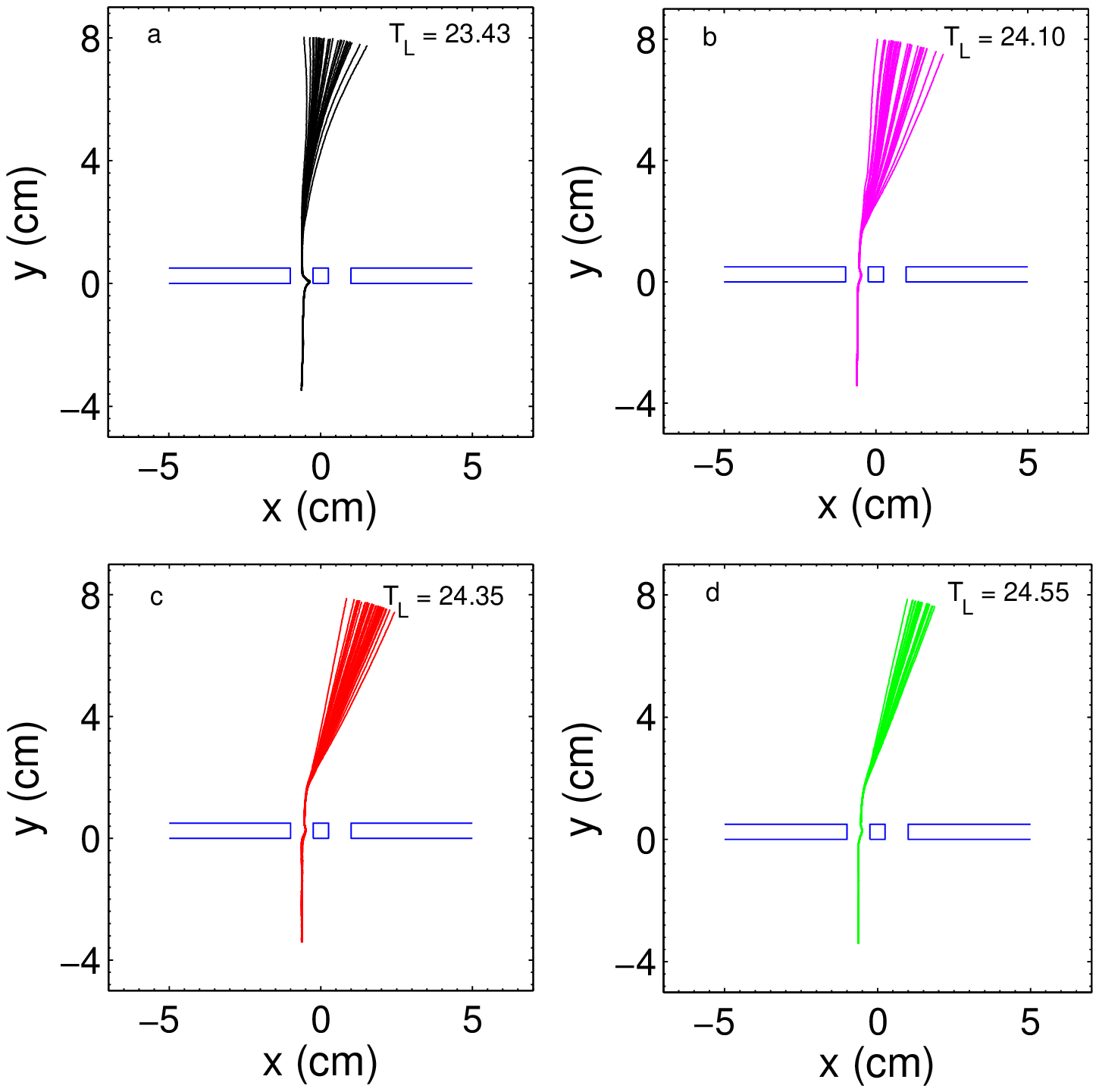}
\caption{(Color online) Track development as the temperature T$_L$ of the glass lid is varied. $x_{imp}$ fixed. a) T$_L$ = 23.43 $^{\circ}{\rm C}$. b) T$_L$ = 24.10 $^{\circ}{\rm C}$. c) T$_L$ = 24.35 $^{\circ}{\rm C}$. d) T$_L$ = 24.55 $^{\circ}{\rm C}$. T$_{oil}$ varies from 25.106-25.135 $^{\circ}{\rm C}$ from a) to d). $\gamma/\gamma_F = 0.996$, v$_{imp}$ = 9.4 mm/s, h$_1$ = 0.64 mm. 7.3 mm double slit.} 
\label{tempstab}
\end{figure}

Here we discuss the influence of the glass temperature. In Fig.~\ref{tempstab} we show the development of a series of tracks where the impact parameter $x_{imp}$ and oil temperature T$_{oil}$ are kept constant while the temperature T$_L$ of the glass lid is varied.
At the lowest temperature of 23.43 $^{\circ}{\rm C}$ the tracks show noticeable bending. As the glass temperature is raised both the bending and the spread of the tracks on the output side become less until at 24.55 $^{\circ}{\rm C}$ we encounter a comparatively narrow bunch of straight tracks. 

\noindent
\section{Appendix D}
\begin{table*}
\begin{tabular}{lcccccc}
Experiment  & V (mm/s) & $\sqrt{2}$d (mm) & $\gamma_F/(\gamma_F - \gamma)$ & MeT$_F$V$_w/\sqrt{2}$d  & {\it l}/$\sqrt{2}$d & R/{\it l} \\
\\
Couder \& Fort \cite{CouderFort,Tadrist}                 & 19   & 20.2 & 5-20 & 0.12-0.5 & 0.8  & 2.5-4.7 \\
Andersen {\it et al.} \cite{Andersen}  & 15   & 14.1 & 21   &  0.5     & 1.17 & 2.82    \\
Pucci {\it et al.} \cite{BushNew}      & 6.8  & 28.3 & 500  &  3.3     & 2.6  & 1.08    \\
7.3 mm double-slit Fig.~\ref{Ang_dist}                    & 8.3  & 16.9 & 250  &  3.1     & 3.2  & 1.67    \\    
7.3 mm double-slit Fig.~\ref{Ang_dist2}                   & 11.0 & 16.9 & 1000 &  16.3    & 4.5  & 1.18    \\
5 mm double-slit Fig.~\ref{Block}                       & 11.1 & 13.7 & 1000 &  20.3    & 5.6  & 1.18   
\end{tabular}
\caption{Overview of experimental conditions used in previous experiments Ref.~\cite{Tadrist}  compared with those of the  present experiment on double-slit barriers. V$_w$ is the droplet velocity, d the distance between slit centers, and R is the radius of the deep section of the container. The path length of the waves $\sqrt{2}$d, the normalized damping time MeT$_F$V$_w/\sqrt{2}$d, and the normalized damping length   {\it l/}$\sqrt{2}$d are calculated using the theory of Tadrist  {\it et al.} \cite{Tadrist}.}
\label{table1}
\end{table*}

Here we present an overview of parameters used for double-slit experiments in Table~\ref{table1}. Also presented are some parameters obtained by using  theoretical criteria derived by Tadrist {\it et al.} \cite{Tadrist} for the observation of true interference in a double-slit experiment. Data for the Couder \& Fort experiment are taken from  Tadrist {\it et al.}. As seen, quite a range of parameters have been investigated over  the years.

Using their nomenclature, the  criteria state that the normalized damping time MeT$_F$V$_w/\sqrt{2}$d, the normalized damping length   {\it l}/$\sqrt{2}$d, and the normalized radius R/{\it l} should all be significantly larger than unity. Here Me is the wave memory,  {\it l} is a damping length, V$_w$ is the droplet velocity, d the distance between slit centers, and R is the radius of the deep section of the container.

As seen, the first two criteria are only satisfied marginally in the experiment by Pucci {\it et al.} \cite{BushNew} and our low memory experiment, but reasonably well in our high memory cases. So why do we not see the interference even here.

The most likely reason is that the theory does not take into account the finite thickness of the barrier. This has two important consequences. As the passage time of the slit is neglected, the normalized damping time becomes much too large. Secondly, the finite barrier thickness effectively prohibits the waves from entering the unused slit when the droplet approaches the slit used. The outcome of this is that the third condition, that the normalized container size R/{\it l} $>>$ 1, never comes into play.


\begin{thebibliography}{99}
\bibitem{unknownT}The spread in the data might be due to a difference in oil temperature as none of these authors supply information on this parameter. We assume that all data are taken at 25 $^{\circ}{\rm C}$.
\bibitem{FitCurve}The data found in the literature can be fitted to h(x) = -1.54 + 1.8 x$^{0.1}$. While arguably other fitting curves than this slowly curving one could be chosen due to the scatter in the data, it does have the additional virtue of fitting our data if our droplet sizes are all increased by about 4\%.
\bibitem{PDF}A MATLAB program is used to generate a smooth curve. The fits are made by re-sampling and smoothing the data. Following a symmetrization using the appropriate symmetry, a final re-sampling and smoothing is done. Resampling filter used is 'pchip' with number of points N of order 10000.  Smoothing filter is 'rloess' or 'rlowess' with a span around 0.05. The actual choice of N, smoothing filter, and span for this, depends on the complication involved. The density of the re-sampled points is then projected onto the angular response axis. The resulting plots are clearly not true probability density functions as the angular response is not a random stochastic variable.
\bibitem{ADXL}The level of sideways oscillation was investigated by placing 4 ADXL accelerometers on the container, shifting their positions and rotating them to remove the effect of cross-axis leakage. As the accelerometers are highly temperature sensitive the temperature stabilization was turned on.
\end{thebibliography}
\end{document}